\documentclass[namedreferences]{SolarPhysics}
\usepackage{spr-sola-addons} 
\usepackage{epsfig}                     
\usepackage{graphicx}                    
\usepackage{color}                       
\usepackage{url}                         
\usepackage{amsfonts}
\usepackage{lineno}

\begin{document}
\begin{article}
\begin{opening}
\title{Roles of Fast-Cyclotron and Alfv\'{e}n-Cyclotron Waves for the Multi-Ion Solar Wind}

\author{Ming.~\surname{Xiong}$^{1}$\sep
        Xing.~\surname{Li}$^{1}$\sep
       }

%

  \institute{$^{1}$ Aberystwyth University, UK
                     email: \url{mxiong@spaceweather.ac.cn} email: \url{xxl@aber.ac.uk}
             }

\begin{abstract}
Using linear Vlasov theory of plasma waves and quasi-linear theory
of resonant wave-particle interaction, the dispersion relations
and the electromagnetic field fluctuations of fast and Alfv\'{e}n
waves are studied for a low-beta multi-ion plasma in the inner
corona. Their probable roles in heating and accelerating the solar
wind {\it via} Landau and cyclotron resonances are quantified. In
this paper, we assume that (1) low-frequency Alfv\'{e}n and fast
waves, emanating from the solar surface, have the same spectral
shape and the same amplitude of power spectral density (PSD); (2)
these waves eventually reach ion cyclotron frequencies due to a
turbulence cascade; (3) kinetic wave-particle interaction powers
the solar wind. The existence of alpha particles in a dominant
proton/electron plasma can trigger linear mode conversion between
oblique fast-whistler and hybrid alpha-proton cyclotron waves. The
fast-cyclotron waves undergo both alpha and proton cyclotron
resonances. The alpha cyclotron resonance in fast-cyclotron waves
is much stronger than that in Alfv\'{e}n-cyclotron waves. For
alpha cyclotron resonance, an oblique fast-cyclotron wave has a
larger left-handed electric field fluctuation, a smaller wave
number, a larger local wave amplitude, and a greater energization
capability than a corresponding Alfv\'{e}n-cyclotron wave at the
same wave propagation angle $\theta$, particularly at $80^\circ <
\theta < 90^\circ$. When Alfv\'{e}n-cyclotron or fast-cyclotron
waves are present, alpha particles are the chief energy recipient.
The transition of preferential energization from alpha particles
to protons may be self-modulated by differential speed and
temperature anisotropy of alpha particles {\it via} the
self-consistently evolving wave-particle interaction. Therefore,
fast-cyclotron waves as a result of linear mode coupling is a
potentially important mechanism for preferential energization of
minor ions in the main acceleration region of the solar wind.
\end{abstract}


\end{opening}

\section{Introduction}
The solar wind is a ubiquitous super-Alfv\'{e}nic stream of
magnetized charged particles from the Sun, consisting of protons,
electrons, and minor ions. According to spectroscopic observations
of extreme ultraviolet emission lines in the inner corona, solar
wind ions exhibit a large kinetic temperature anisotropy, and
minor ions are likely to be preferentially heated and accelerated
\cite{Kohl1998,Li1998,Cranmer1999}. In the interplanetary space,
minor ions have mass-proportional temperatures and similar bulk
flow speed, whose drifting speeds relative to protons are at an
order of local Alfv\'{e}n speed
\cite{Marsch1982,Neugebauer1996,Goldstein2000}. And proton
temperature in the direction parallel to the background
interplanetary magnetic field is lower than that in the
perpendicular direction. Double components of tenuous fast beam
and dense core frequently occur in the velocity distribution
function (VDF) of protons, with a differential speed of 1.2 -- 1.3
Alfv\'{e}n speed \cite{Marsch1982}. The combination of these
remote-sensing and interplanetary {\it in-situ} measurements
substantiates that some energy should be continuously deposited
into the solar wind ions {\it via} some physical mechanisms, and
thus power the hot collisionless outflow from the Sun.

The solar wind has been an outstanding puzzle for its heating and
acceleration since the theoretical prediction \cite{Parker1958}
and observation verification \cite{Neugebauer1962}. A fraction of
the mechanical energy in the Sun's internal convective motions
must be converted into thermal energy above the photosphere. Ion
cyclotron waves are generally considered as a promising mechanism
that can preferentially heat and accelerate minor ions. Ion
cyclotron frequencies in the extended corona vary from 10 to
$10^4$ Hz, whereas observed frequencies of pronounced oscillation
on the solar surface are typically $\approx 0.01$ Hz. How strong
enough ion cyclotron waves could be generated is still debated. In
terms of the origins of ion cyclotron waves, there are two main
viewpoints of base and secondary generations. For the base
generation, high-frequency Alfv\'{e}n waves are directly launched
during small-scale magnetic reconnection of micro-flares in
rapidly evolving super-granular networks, then propagate up
through the corona, gradually reach cyclotron frequencies of
various ions, and are finally damped over a very short distance
\cite{Axford1992}. For the secondary generation, high-frequency
ion cyclotron waves are gradually converted from low-frequency
Alfv\'{e}n waves in the corona, and such conversion process is
probably from a direct turbulence cascade
\cite{li2001,Markovskii2010}, local nonlinear kinetic
instabilities \cite{Markovskii2004}, and Debye-scale electron
phase space holes in electron Langmuir turbulence
\cite{Cranmer2003}. In this paper, we assume that (1) the plasma
turbulence cascade is the energy source of the solar wind and (2)
the solar wind heating is due to Landau and cyclotron resonances
with a continuous turbulence spectrum.

The solar wind turbulence is an ensemble of fluctuations with
random phases and a broad range of wave vectors. Such fluctuations
are manifested in magnetic field $\delta \mathbf{B}$, electric
field $\delta \mathbf{E}$, plasma bulk flow velocity $\delta
\mathbf{V}$, and so on. The fluctuation energy is usually
transferred through a forward cascade. Namely, the energy is
injected from a preexisting population of magnetohydrodynamic
(MHD) waves at very low frequencies $f$ and wave numbers $k$,
nonlinearly transported through successively shorter wavelengths
{\it via} wave-wave interaction, and eventually dissipated at ion
kinetic scales {\it via} collisionless wave-particle interaction.
The evolution of $\delta \mathbf{B}$ and $\delta \mathbf{V}$ in
the presence of a background magnetic field $\mathbf{B_0}$ can be
represented in the Elsasser variables $\mathbf{Z}^\pm = \delta
\mathbf{V} \pm \delta \mathbf{B}/\sqrt{4\,\pi \rho_0}$ as follows:
\begin{equation}
\frac{\partial \mathbf{Z}^\pm}{\partial t} \mp (v_{\rm A} \cdot
\nabla) \mathbf{Z}^\pm + (\mathbf{Z}^\mp \cdot \nabla)
\mathbf{Z}^\pm = - \nabla P + F, \label{Equ:Elsasser}
\end{equation}

\begin{eqnarray}
    \left\{ \begin{array}{ll}
    k_\parallel \,B_0 \gg k_\perp b_\lambda \, , & \quad \mbox{weak turbulence,} \\
    k_\parallel \,B_0 \approx k_\perp b_\lambda \, , & \quad \mbox{strong turbulence.}
    \end{array} \right. \label{Equ:weak-strong}
\end{eqnarray}
Here $v_{\rm A}=B_0 / \sqrt{4\pi \rho_0}$ is the Alfv\'{e}n speed,
$\rho_0$ is the mass density, $P$ is the pressure that is
determined from the incompressibility condition of $\nabla \cdot
\mathbf{Z}^\pm = 0$, $F$ is a large-scale forcing, $k_\parallel$
and $k_\perp$ are parallel and perpendicular wave numbers, and
$b_\lambda$ ($\ll B_0$) is oscillation amplitude at the scale
$\lambda \approx 1/k_\perp$. $\mathbf{Z}^\pm$ is imbalanced at a
macro scale, as supported by nonzero cross-helicity in terms of
$\delta \mathbf{V} \cdot \delta \mathbf{B}$
\cite{Goldreich1995,He2011}. There is more power in Alfv\'{e}nic
fluctuations travelling away from the Sun than towards it. Even if
balanced overall, MHD turbulence is always locally imbalanced in
creating patches of positive and negative cross-helicities. The
conservation of cross-helicity results in a hierarchical structure
of MHD turbulence. In the corona, turbulence amplitude is
constrained by interplanetary scintillation (IPS) observations of
electron density spectra \cite{Cole1978}. Interplanetary {\it in
situ} measurements of turbulence $\delta \mathbf{B}$ as a function
of $f$, $k_\parallel$, and $k_\perp$ have shown a power law
spectrum of magnetic field fluctuation, and a large anisotropy in
favor of quasi-perpendicular wave propagation
\cite{Matthaeus1990,Leamon1998,Dasso2005,Gary2009,MacBride2010}:
\begin{equation}
|\delta \mathbf{B}(f)|^2 \approx f^{-\alpha}, \label{Equ:PSD}
\end{equation}
\begin{equation}
|\delta \mathbf{B}(k_\parallel)|^2 = \displaystyle \sum_{k_\perp}
|\delta \mathbf{B}(\mathbf{k})|^2 \approx
k_\parallel^{-\alpha_\parallel}, 
\label{Equ:PSD1}
\end{equation}
\begin{equation}
|\delta \mathbf{B}(k_\perp)|^2 = \displaystyle \sum_{k_\parallel}
|\delta \mathbf{B}(\mathbf{k})|^2 \approx k_\perp^{-\alpha_\perp},
\label{Equ:PSD2}
\end{equation}
\begin{equation}
\alpha_\parallel > \alpha_\perp \, .
\end{equation}
The solar wind turbulence in an inertial range of $10^{-4} \leq f
\leq 0.2 $ Hz is similar to the classical Kolmogorov picture of
fluid turbulence, and the spectral index $\alpha$ in Equation
(\ref{Equ:PSD}) is approximately $5/3$
\cite{Leamon1998,Horbury2005}. Meanwhile, the power spectral
density  is anisotropic with respect to $\mathbf{B_0}$, as
indicated from $\alpha_\parallel \approx 2$ in Equation
(\ref{Equ:PSD1}) and $\alpha_\perp \approx 1.6$ in Equation
(\ref{Equ:PSD2}) \cite{Gary2009}. Moreover, within the inertial
range, the Elsasser variables $\mathbf{Z}^\pm$ were also found to
be anisotropic and have two distinct scaling subranges
\cite{Wick2011}. A transition between weak and strong turbulence
regimes (Equation (\ref{Equ:weak-strong})) is feasible by changing
a large-scale forcing $F$ (Equation (\ref{Equ:Elsasser})), as
demonstrated from numerical simulation of anisotropic
incompressible MHD turbulence \cite{Perez2008}. The nature of the
MHD turbulence cascade lies in the interaction of wave packets
moving with Alfv\'{e}n velocities. In the framework of the weak
turbulence scenario, interacting wave packets are very large such
that the original wave packets decay before they pass through each
other. Because the auto-correlation and cascade time scales are
always of the same order of magnitude in the weak turbulence
theory, any model that implies a large number of collisions among
wave packets for an efficient energy cascade should belong to the
strong turbulence theory \cite{Gogoberidze2009}. In the strong
turbulence, interaction is strong, resonance conditions are not in
force, hence a parallel cascade is possible \cite{Perez2008}. The
forward cascade of $\delta \mathbf{B}$ could be modelled as a
combination of advection and diffusion in the wave number
$\mathbf{k}$ space \cite{Cranmer2003}. Equations (\ref{Equ:PSD1})
and (\ref{Equ:PSD2}) for turbulence anisotropy in the inertial
range could be derived by a conjecture of dynamical alignment that
polarizations of $\delta \mathbf{B}$ and $\delta \mathbf{V}$ tend
to align during the turbulence cascade towards successively
smaller scales \cite{Goldreich1995}. A good fraction of the power
spectral density (PSD) observations is well interpreted by the
current turbulence theories.

The turbulence in the dissipation range is poorly understood.
Spectral properties in the dissipation range almost certainly
depend upon the details of the damping, though those in the
inertial range are relatively robust and independent of detailed
turbulence cascade processes. The observed wave number at a
breakpoint, which separates the inertial and dissipation ranges,
may scale as inverse proton inertial length $k \,c/\omega_{\rm pp}
\approx 1$ \cite{Leamon1998,Smith2001} or cyclotron radius $k \,
v_{\rm p} / \Omega_{\rm p} \approx 1$ \cite{Bale2005}, with
$\omega_{\rm pp}$, $\Omega_{\rm p}$, and $v_{\rm p}$ denoting
proton plasma frequency, proton cyclotron frequency, and proton
thermal speed respectively. Above the inertial range, the PSD has
a distinct breakpoint at $0.2 < f < 0.5$ Hz; At higher frequencies
of $f \ge 0.5$ Hz, the PSD is steeper with a broader range of $2
\leq \alpha \leq 4$ in Equation (\ref{Equ:PSD})
\cite{Leamon1998,Gary2009}. For instance,
\inlinecite{Sahraoui2009} found that a PSD above the inertial
range consists of two distinct regimes with successively larger
values of $\alpha$ in Equation (\ref{Equ:PSD}); $\alpha$ is about
2.5 from 0.4 to 35 Hz for a dispersion cascade and 4 from 35 to
100 Hz due to strong wave damping. Furthermore, it is still
unclear what are the constituent wave modes of the solar wind
turbulence. The macro-scale fluctuations in the solar wind and
corona are Alfv\'{e}nic in nature \cite{Barnes1974}. In the MHD
regime, Alfv\'{e}n waves can be converted into fast waves by
density fluctuations, and vice versa through resonant three-wave
interaction \cite{Chandran2008}. At the kinetic scale, highly
oblique whistler waves exist according to the observed properties
of large magnetic compressibility in the solar wind
\cite{Gary2009}. Kinetic Alfv\'{e}n waves are identified from the
observations of largely enhanced electric fluctuation spectrum and
super-Alfv\'{e}nic wave phase speed at scales of ion thermal
gyro-radius \cite{Leamon1998,Bale2005}. Parallel propagating
Alfv\'{e}n-cyclotron waves are possibly present as evidenced by
the measured angle distribution of magnetic helicity
\cite{He2011}. These observations have revealed the importance of
quasi-perpendicularly propagating waves in the heating of the
solar wind.

Increasing attention has been paid to oblique propagation owing to
the natural occurrence of oblique waves from refractions in the
inhomogeneous solar wind, production by density gradients, and
generation from MHD turbulence
\cite{li2001,Gary2004a,Osmane2010,Li2010b,Markovskii2010}. Indeed,
for the Alfv\'{e}n-cyclotron wave branch, parallel propagation has
its inherent difficulty to reconcile both the heating and the
acceleration of the solar wind \cite{Hollweg2002b}. The nature of
obliquely propagating waves is subject to debates
\cite{Gary2009,Podesta2010}. According to interplanetary
measurements, these constituent waves may be kinetic Alfv\'{e}n
waves \cite{Bale2005, Sahraoui2010} or fast-whistler waves
\cite{Gary2009}. In a warm plasma such as the interplanetary solar
wind, oblique Alfv\'{e}n waves can generate a fast proton beam
through Landau resonance \cite{Li2010b,Osmane2010}. In a cold
proton/electron plasma, quasi-perpendicular fast-whistler waves
break up into Bernstein modes near the first few harmonics of
proton cyclotron frequency \cite{li2001,Markovskii2010}. Due to
the existence of minor ions in a solar corona-like plasma, oblique
fast-whistler waves are linearly coupled with hybrid cyclotron
waves \cite{li2001}. Depending on the abundance and relative speed
with respect to protons, minor ions may change wave dispersion
relation dramatically, induce a transition of wave polarization
from right- to left-handed sense, and can absorb more energy from
oblique fast-cyclotron waves than Alfv\'{e}n-cyclotron waves for
intermediate angles $30^\circ$ -- $50^\circ$ of wave propagation
\cite{li2001}. However, Alfv\'{e}n-cyclotron and fast-cyclotron
waves in the inner corona have few empirical constraints for their
exact properties such as generation mechanisms, propagation modes,
and power levels. With favorable plasma parameters, both
Alfv\'{e}n and fast waves can be potential candidates for
preferential energization of minor ions in the solar wind. On the
basis of current understanding, both fast-whistler and
Alfv\'{e}n-cyclotron waves at both quasi-parallel and
quasi-perpendicular propagation angles probably contribute to
kinetic-scale turbulence in the solar wind \cite{Gary2009,He2011}.

In this paper, both Alfv\'{e}n and fast waves are investigated to
explore their roles in energizing the low-beta multi-ion solar
wind plasma in the inner corona. As a continuation to the work of
\inlinecite{li2001}, the linear mode conversion between
fast-whistler and hybrid cyclotron waves is further quantified by
a detailed parametric study of various plasma parameters. A
quasi-linear theory of resonant wave-particle interaction
\cite{Marsch2001} is used, and we are able to study the parallel
and perpendicular heating of solar wind species due to Landau and
cyclotron resonances. We give our model in Section
\ref{Sec:Method}, and describe wave dispersion relations and
accompanying electromagnetic field oscillations in Section
\ref{Sec:dispersion}. Then the roles of wave propagation angle,
minor ion presence, relative flow of minor ions, and plasma
temperature are successively analyzed in Sections \ref{Sec:angle},
\ref{Sec:minor-ion}, \ref{Sec:dif-V}, and \ref{Sec:temperature}.
We discuss the ensuing nonlinear stage of wave-particle
interaction in Section \ref{Sec:PSD}, and summarize our results in
Section \ref{Sec:Conclusion}.

\section{Methods}\label{Sec:Method}
To evaluate the energization effect of kinetic wave-particle
interaction on the multi-species solar wind, two methods are
sequentially combined in this paper. First, using linear Vlasov
theory \cite{Stix1992,li2001}, the properties of Alfv\'{e}n and
fast waves such as their dispersion relations and electromagnetic
polarizations are studied. Second, using a quasi-linear theory of
resonant wave-particle interaction \cite{Marsch2001}, the
previously derived wave properties are adopted to calculate the
acceleration and heating rates of different plasma species. With
the two-step approach, the wave-particle interaction can be
quantified to demonstrate the most viable and efficient way for
the solar wind energization.

Since we are interested in the main solar wind acceleration
region, typical plasma conditions in the inner corona are chosen
in our analyses. Given a background magnetic field $\mathbf{B_0}$,
a wave vector $\mathbf{k}$ is decomposed into a parallel component
$k_\parallel$ and a perpendicular one $k_\perp$. The propagation
angle of a wave $\theta = \arctan (k_\perp / k_\parallel)$ is
defined with respect to $\mathbf{B_0}$. Our Cartesian system of
$(x, y, z)$ coordinates is constructed with $\mathbf{B_0}$ along
the $z$-axis and $\mathbf{k}$ within the $xz$-plane. The symbols
and corresponding denotations used in this paper are summarized in
Table \ref{Tab1}. In this paper, only electrons $\rm e$, protons
$\rm p$, and alpha particles $\alpha$ are considered in the solar
wind. For simplicity, VDF of each species is approximated to be
bi-Maxwellian. The parallel (perpendicular) thermal speed
$v_{\parallel j}$ ($v_{\perp j}$) and gyrotropic VDF
$f_j(v_\parallel, v_\perp)$ are expressed as follows:
\begin{eqnarray}
 v_{\parallel j}  &=& \sqrt{2 \, k_{\rm B} T_{\parallel j} / m_j} \,\, ,\\
 v_{\perp j} &=& \sqrt{2 \, k_{\rm B} T_{\perp j} / m_j} \,\, ,\\
 f_j(\mathbf{v}) &=& f_j(v_\parallel, v_\perp) = \frac{1}{\pi^{3/2} \, v_{\parallel j}\, v^2_{\perp j}} \exp
\left( - \frac{(v_\parallel - U_{\parallel j})^2}{v_{\parallel
j}^2} - \frac{v_\perp^2}{v_{\perp j}^2} \right). \label{Equ:VDF}
\end{eqnarray}

The dispersion relation $\omega=\omega(\mathbf{k})$ and its
inherent electromagnetic field ($\delta \mathbf{E}$, $\delta
\mathbf{B}$) of a plasma wave are calculated using linear Vlasov
theory \cite{Stix1992,li2001}. The complex wave frequency
$\mathbf{\omega}(\mathbf{k}) = \omega_{\rm r}(\mathbf{k}) + i
\,\omega_{\rm i} (\mathbf{k})$, $\delta \mathbf{B}(\mathbf{k})$,
and $\delta \mathbf{E}(\mathbf{k})$ are found as a function of
wave vector $\mathbf{k}$. The damping rate $|\omega_{\rm i}|$
generally increases with wave number $k$. \inlinecite{Gary1993}
suggested that heavily attenuated waves are described by a rough
approximation:
\begin{eqnarray}
\omega_{\rm i} (\mathbf{k}) < - \frac{\omega_{\rm r}
(\mathbf{k})}{2 \, \pi} \,\, . \label{Equ:Gary-threshold}
\end{eqnarray}
and that only weakly damped waves can persist. Given $\delta
\mathbf{E}(\mathbf{k})$, polarization $P$ and unimodular electric
vector $\mathbf{e}_E(\mathbf{k})$ are given by
\begin{eqnarray}
 P &=& i \frac{\delta \mathbf{E}_x}{\delta \mathbf{E}_y}
\frac{\omega_{\rm r}}{|\omega_{\rm r}|} \,\, ,\\
\mathbf{e}_E(\mathbf{k})&=& \frac{\mathbf{\delta
E}(\mathbf{k})}{|\mathbf{\delta E}(\mathbf{k})|} \,\, .
\end{eqnarray}
Consequently, the magnetic field fluctuation $\mathbf{\delta
B}(\mathbf{k})$ and its power spectrum $|\delta
\mathbf{B}(\mathbf{k})|^2$ are
\begin{eqnarray}
 \delta \mathbf{B}(\mathbf{k}) &=& \frac{1}{\omega(\mathbf{k})} \,
\mathbf{k} \times \mathbf{\delta E}(\mathbf{k})\,\, , \\
 |\delta \mathbf{B}(\mathbf{k})|^2 &=& \left(\frac{k}{\omega_{\rm r}(\mathbf{k})} \right)^2 (1- |\mathbf{e}_k \cdot
 \mathbf{e}_E(\mathbf{k})|^2) \,\, |\delta
 \mathbf{E}(\mathbf{k})|^2  \label{Equ:mag2}\,\, , \\
\mathbf{e}_k&=& \frac{\mathbf{k}}{|\mathbf{k}|} =
\frac{\mathbf{k}}{k}\,\, .
\end{eqnarray}

The wave-particle interaction {\it via} Landau and cyclotron
resonances is efficient in transferring wave energy to particles
in a collisionless plasma. With the introduction of
Doppler-shifted frequency $\omega'(\mathbf{k}) = \omega_{\rm r}
(\mathbf{k}) - k_\parallel \, U_{\parallel j}$, the $n$th-order
resonant parallel speed in the bulk frame of species $j$ is
defined as:
\begin{equation}
w_j(\mathbf{k}, n) = \frac{\omega'(\mathbf{k}) -
n\,\Omega_j}{k_\parallel} = \frac{\omega_r (\mathbf{k}) -
k_\parallel \, U_{\parallel j} - n\,\Omega_j}{k_\parallel} \,\, .
\label{Equ:Doppler}
\end{equation}
Then, the resonant factor $\xi_j$ for species $j$ is defined as:
\begin{equation}
  \xi_j(\mathbf{k}, n) = \frac{w_j(\mathbf{k}, n)}{v_{\parallel j}} = \frac{\omega_r (\mathbf{k}) - k_\parallel \, U_j - n \,\Omega_j}{k_\parallel
  \, v_{\parallel j}}, \quad n=0, \pm 1,\pm 2,... \,\, .
\end{equation}
Here $n=0$ corresponds to the Landau resonance, $n \neq 0$ the
cyclotron resonance. Only the Landau resonance $\xi^0$ and the
first-order cyclotron resonance $\xi^+$ may appear for the plasma
parameters prescribed in Tables \ref{Tab2} and \ref{Tab3}.

The acceleration and heating rates of a species due to any wave
mode were formulated by \inlinecite{Marsch2001} using a
quasi-linear theory, and are applied in this paper. In the
quasi-linear theory, different waves are simply considered to be
linearly superposed without any interference. Specifically, for
species $j$, its parallel acceleration rate $a_{\parallel j}$,
parallel heating rate $Q_{\parallel j}$, and perpendicular heating
rate $Q_{\perp j}$ are integrated over various wave modes $M$ and
all resonance orders $n$, and are listed below:
\begin{equation}
 \begin{array}{l}
 \left( \begin{array}{c}
   \frac{\partial}{\partial t} U_{\parallel j}\\
   \frac{\partial}{\partial t} v_{\parallel j}^2\\
   \frac{\partial} {\partial t} v_{\perp j}^2\\
 \end{array}\right) =
 \displaystyle \sum_M \int^{+ \infty}_{- \infty} d^3k \, \frac{
 \displaystyle  |\delta \mathbf{B}(\mathbf{k})|^2}{8 \, \pi}
\left( \begin{array}{c}
   a_{\parallel j} \, / \, m_j \\
   Q_{\parallel j} \, / \, m_j \\
   Q_{\perp j} \, / \, m_j \\
 \end{array}\right)
 \end{array},\label{Equ:Marsch}
\end{equation}

\begin{equation}
\hspace*{-9mm}
 \begin{array}{l}
 \left( \begin{array}{c}
   a_{\parallel j} \, / \, m_j \\
   Q_{\parallel j} \, / \, m_j \\
   Q_{\perp j} \, / \, m_j \\
 \end{array}\right)
 = \displaystyle \sum_{n=-\infty}^{+\infty}
 \left(\frac{1}{2 \pi}\right)^3  \left( \frac{\Omega_j}{k} \right)^2
 \frac{1}{1- |\mathbf{e_k} \cdot
 \mathbf{e}_E(\mathbf{k})|^2} \,
   R_j(\mathbf{k},n) \left(  \begin{array}{c}
   k_\parallel \\
   2 \,k_\parallel \, w_j(\mathbf{k},n) \\
   n \,\Omega_j \\
  \end{array}\right)
  \end{array}. \label{Equ:aQ}
\end{equation}

Herein, the resonance function $R_j(\mathbf{k},n)$ gives energy
absorption of species $j$ from a wave mode $|\delta
\mathbf{B}(\mathbf{k})|^2$. An electric field oscillation could be
factorized into a left-handed polarization and a right-handed one.
The factorized components of unimodular electric field
$\mathbf{e}_E(\mathbf{k})$ and total electric field $\delta
\mathbf{E}(\mathbf{k})$ are expressed as
\begin{eqnarray}
\mathbf{e}^\pm_{E}(\mathbf{k}) &=& \mathbf{e}_{Ex}(\mathbf{k}) \pm
i \, \mathbf{e}_{Ey}(\mathbf{k})\,\, , \label{Equ:e-pm} \\
\delta \mathbf{E}^\pm(\mathbf{k}) &=& \delta
\mathbf{E}_x(\mathbf{k}) \pm i \, \delta \mathbf{E}_y(\mathbf{k})
\,\, . \label{Equ:E-pm}
\end{eqnarray}
In Equations (\ref{Equ:e-pm}) and (\ref{Equ:E-pm}), subscripts $+$
and $-$ denote left- and right-handed polarizations, respectively.
The transverse wave number $k_\perp$ is compared with the ion
cyclotron radius of thermal motions $v_{\perp j} / \Omega_j$, and
hence described by a parameter $\lambda_j$.
\begin{equation}
\lambda_j = \frac{1}{2} \left( \frac{k_\perp \, v_{\perp
j}}{\Omega_j} \right)^2 \,\,.
\end{equation}
Further, the resonance function $R_j(\mathbf{k},n)$ is defined as
\begin{eqnarray}
R_j(\mathbf{k},n) &=& 2 \, N_j(\mathbf{k},n) \, \exp(-\lambda_j)\,
[a_-
I_{n-1}(\lambda_j) + a_+ I_{n+1}(\lambda_j) + a_0 I_n(\lambda_j)]\,\,, \label{Equ:Rjkn}\\
N_j(\mathbf{k},n) &=& \sqrt{\pi} \,
\frac{k_\parallel}{|k_\parallel|} \exp[-\xi_j^2(\mathbf{k},n)]
\times \left\{ {\xi_j(\mathbf{k},n) \,\frac{T_{\perp j}}{T_{j
\parallel}} + n\, \frac{\Omega_j}{k_\parallel \, v_{\parallel j}}}
\right\}\,\,, \label{Equ:Njkn}\\
a_0(\mathbf{k},n) &=& 2 \left( \frac{\omega(\mathbf{k}) - n\,
\Omega_j}{k_\parallel \, v_{\perp j}} \right)^2
|\mathbf{e}_{Ez}|^2 -  2 \mbox{ Im}\{\mathbf{e}^*_{Ey}
\mathbf{e}_{Ez}\} \frac{k_\perp}{k_\parallel}
\frac{\omega(\mathbf{k}) - n \,\Omega_j}{\Omega_j} + 2
\,\lambda_j \, |\mathbf{e}_{Ey}|^2 \,\,, \nonumber\\ \\
a_\mp(\mathbf{k},n) &=& \pm \frac{n}{2} \, |\mathbf{e}^\pm_E|^2
\pm \mbox{ Re}\{\mathbf{e}^*_{Ez} \mathbf{e}^\pm_E\}
\frac{k_\perp}{k_\parallel} \frac{\omega(\mathbf{k}) - n \,
\Omega_j}{\Omega_j} \mp
 \lambda_j
\mbox{ Im}\{\mathbf{e}^*_{Ey} \mathbf{e}^\pm_E\} \,\,.
\end{eqnarray}
Here $I$ is a modified Bessel function. For parallel propagation,
the above form of $R_j(\mathbf{k},n)$ can be largely simplified as
\begin{equation}
R_j(0,k_\parallel, n) = N_j(0, k_\parallel, n) \left[
\delta_{n,1}\, |\mathbf{e}_E^+|^2 + \delta_{n,-1}\,
|\mathbf{e}_E^-|^2 + 4 \, \delta_{n,0} \,|\mathbf{e}_{Ez}|^2
\left( \frac{\omega(k_\parallel)}{k_\parallel \, v_{\perp j}}
\right)^2 \right] \,\,.
\end{equation}
Note that the resonance function $R_j(\mathbf{k},n)$ essentially
depends on the resonance factor $\xi_j$ and the square of the
re-scaled perpendicular wave number $\lambda_j = (k_\perp v_{\rm
A} / \Omega_j)^2/2$ as exponential functions of $\exp( - \xi_j^2)$
in $N_j(\mathbf{k},n)$ (Equation (\ref{Equ:Njkn})) and $\exp(-
\lambda_j)$ in $R_j(\mathbf{k},n)$ (Equation (\ref{Equ:Rjkn}))
respectively. As $\xi_j$ or $\lambda_j$ is increased, $\exp( -
\xi_j^2)$ or $\exp(- \lambda_j)$ dramatically declines to zero,
and energization effect quickly becomes negligible.

We assume that plasma turbulence $|\delta
\mathbf{B}(\mathbf{k})|^2$ is the ultimate energy source of the
solar wind. For simplicity, energization contributions of
Alfv\'{e}n and fast waves are individually evaluated in terms of
the energization rates $a_\parallel$, $Q_\parallel$, and $Q_\perp$
in Equation (\ref{Equ:aQ}). Both Alfv\'{e}n and fast waves are
assumed to have the same PSD $|\delta \mathbf{B}(\mathbf{k})|^2$
at the same wave vector $\mathbf{k}$. A specific form for $|\delta
\mathbf{B} (\mathbf{k})|^2$ in Equation (\ref{Equ:Marsch}) is
beyond the scope of this paper. Initial prescriptions of various
plasma parameters are given in Tables \ref{Tab2} and \ref{Tab3}.
Then the values of $a_\parallel$, $Q_\parallel$, and $Q_\perp$
(Equation (\ref{Equ:aQ})) normalized to $|\delta \mathbf{B}
(\mathbf{k})|^2$ (Equation (\ref{Equ:Marsch})) are derived by a
combined two-step approach \cite{li2001,Marsch2001}. Here a
detailed parametric study is conducted to analyze how cascaded
turbulence energy is repartitioned among multiple constituent
plasma species.

\section{Dispersion Relation and Electromagnetic Field Oscillation}\label{Sec:dispersion}
\begin{figure}
\includegraphics[width=.99\textwidth]{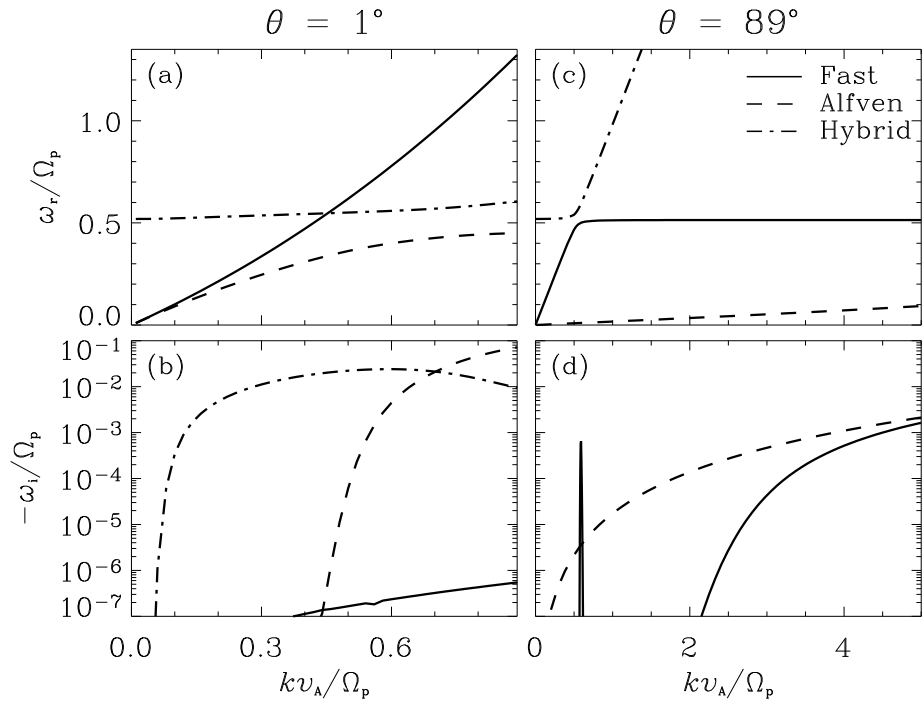}
\caption{Dispersion relations of Alfv\'{e}n, fast, and hybrid
alpha-proton cyclotron waves at (a,b) $\theta=1^\circ$ and (c,d)
$\theta=89^\circ$ in a proton/alpha/electron plasma. Note that
linear mode coupling between hybrid and fast waves happens in
panel (c).} \label{wk}
\end{figure}

\begin{figure}
\includegraphics[width=.7\textwidth]{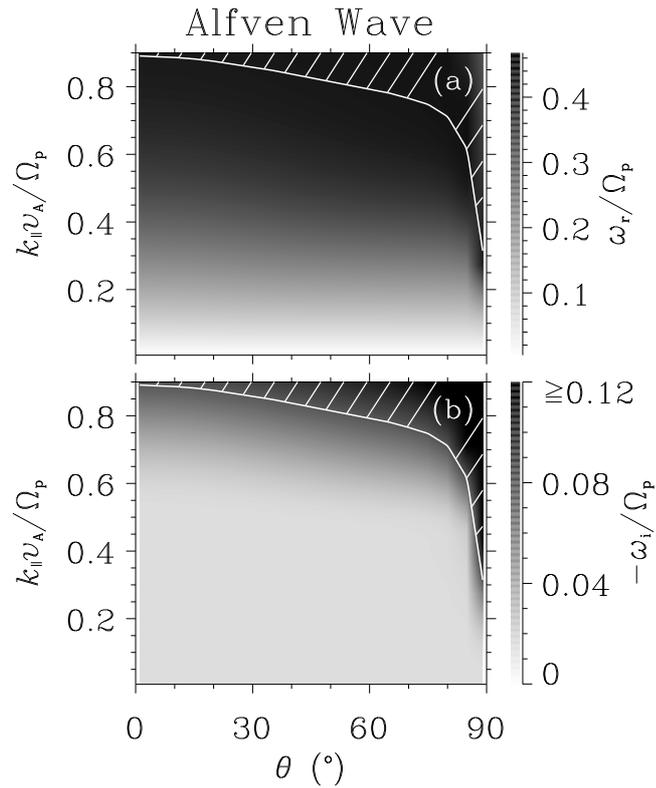}
\caption{For Alfv\'{e}n waves, the dependence of complex frequency
($\omega=\omega_{\rm r} + i \, \omega_{\rm i}$) on the parallel
wave number $k_\parallel$ and propagation angle $\theta$. Heavily
damped waves satisfying $\omega_{\rm i}(\mathbf{k}) < -
\frac{\omega_{\rm r}(\mathbf{k})}{2 \, \pi}$ are indicated by a
hatched region.}\label{wk-Alfven}
\end{figure}

\begin{figure}
\includegraphics[width=.7\textwidth]{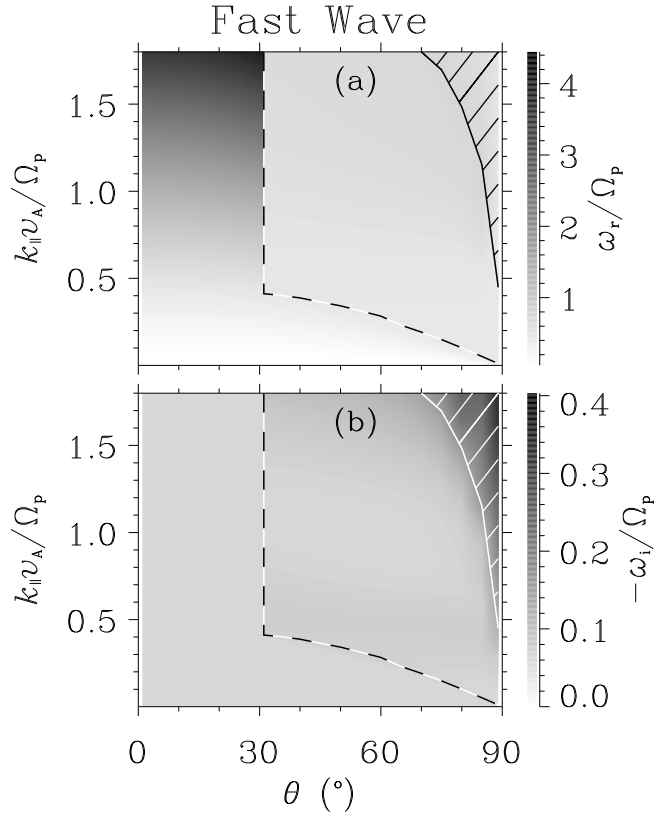}
\caption{For Fast waves, the dependence of complex frequency
($\omega=\omega_{\rm r} + i \, \omega_{\rm i}$) on the parallel
wave number $k_\parallel$ and propagation angle $\theta$. Linear
mode conversion occurs for $\theta \ge 31^\circ$, with its
position in the $k_\parallel-\theta$ domain denoted by a
black-and-white line.}\label{wk-Fast}
\end{figure}

Dispersions of Alfv\'{e}n and fast waves in a
proton/alpha/electron plasma are illustrated for various
propagation angles $\theta$ in Figures \ref{wk}, \ref{wk-Alfven},
and \ref{wk-Fast}. In the presence of alpha particles, an
Alfv\'{e}n-cyclotron wave branch of $0<\omega_{\rm r} <
\Omega_{\rm p}$ is generally split into two parts near the
cyclotron frequency of alpha particles $\Omega_\alpha$. The lower
branch of $0<\omega_{\rm r} < \Omega_\alpha$ corresponds to
Alfv\'{e}nic alpha cyclotron waves, and the upper one of
$\Omega_\alpha <\omega_{\rm r} < \Omega_{\rm p}$ refers to hybrid
alpha-proton cyclotron waves. Obviously, the alpha cyclotron
resonances occur at much smaller wave numbers for the hybrid
cyclotron waves than the Alfv\'{e}n-cyclotron waves. Though having
not been measured experimentally, kinetic hybrid waves with
$\omega_{\rm r} \approx \Omega_\alpha$ and $k \approx 0$ may be
locally generated in a cold multi-ion plasma as a result of linear
mode coupling with fast-whistler waves \cite{li2001}. Avoiding a
crossing between the $\omega - k$ dispersion curves of hybrid
ion-ion cyclotron and fast-whistler waves, fast-whistler waves are
possibly transformed into fast-cyclotron waves. By comparing
Figure \ref{wk}a with \ref{wk}c, one can see that the dispersion
of oblique fast waves is dramatically changed by the presence of
alpha particles in a dominant electron/proton plasma.
Fast-whistler and alpha-proton cyclotron waves exchange their
identities at alpha cyclotron frequency $\omega_{\rm r} =
\Omega_\alpha = 0.5 \, \Omega_{\rm p}$. Such linear mode
conversion in Figure \ref{wk}c occurs when fast waves are oblique
enough, {\it i.e.} the propagation angle $\theta \ge 31^\circ$ in
this paper (Figure \ref{wk-Fast}a). Once mode conversion happens,
the conversion point in the dispersion curve is basically at one
identical point of $\omega_{\rm r} / \,\Omega_{\rm p}=0.5 $ and $k
\, v_{\rm A} /\, \Omega_{\rm p} =0.5$, irrespective of wave
direction $\theta$. At a sufficiently long wavelength,
fluctuations of both Alfv\'{e}n and fast waves are essentially
undamped. For ion cyclotron waves, as the parallel wave number
$k_\parallel$ reaches a characteristic dissipation value $k_{\rm
d}$, relevant ions are in cyclotron resonance and the damping
begins abruptly. In a proton/electron plasma of $0.001 \leq
\beta_{\rm p} \leq 0.1$, $k_{\rm d} \approx \, \Omega_{\rm p} /
v_{\rm A}$ for the proton cyclotron resonance was given by
\inlinecite{Gary2004b}; In a cold alpha/proton/electron plasma,
$k_{\rm d} \approx \, \Omega_\alpha / v_{\rm A} = 0.5 \,
\Omega_{\rm p} / v_{\rm A}$ for the alpha cyclotron resonance is
found in this paper. Specifically, for nearly parallel Alfv\'{e}n
waves of $\theta=1^\circ$, the damping rate $|\omega_{\rm i} /
\Omega_{\rm p}|$ is increased from $2\times 10^{-7}$ at $k \,
v_{\rm A} /\, \Omega_{\rm p} = 0.45$ to $4\times 10^{-3}$ at $k \,
v_{\rm A} / \,\Omega_{\rm p} = 0.6$ (Figure \ref{wk}b). Meanwhile,
damping of fast-whistler waves is only pronounced at harmonic ion
cyclotron frequencies for oblique propagation \cite{li2001}.
However, due to linear mode conversion, fast-cyclotron waves will
experience a dramatic increase of damping rate $|\omega_{\rm i}|$,
as shown in Figure \ref{wk}d. Heavily damped waves disappear
swiftly, as described in Equation (\ref{Equ:Gary-threshold})
\cite{Gary1993}. Approaching ion cyclotron resonance frequency,
smaller parallel wave numbers $k_\parallel$ and larger total wave
numbers $k$ are found as $\theta$ increases to be more oblique,
especially at $\theta \approx 90^\circ$ (Figures \ref{wk-Alfven}
and \ref{wk-Fast}). When a turbulence cascade is present,
Alfv\'{e}n-cyclotron, fast-whistler, and fast-cyclotron waves may
continuously transfer wave energy from very low frequencies along
their respective dispersion curves, and ultimately become
constituent modes of kinetic turbulence at ion cyclotron
frequencies.

\begin{figure}
\includegraphics[width=.99\textwidth]{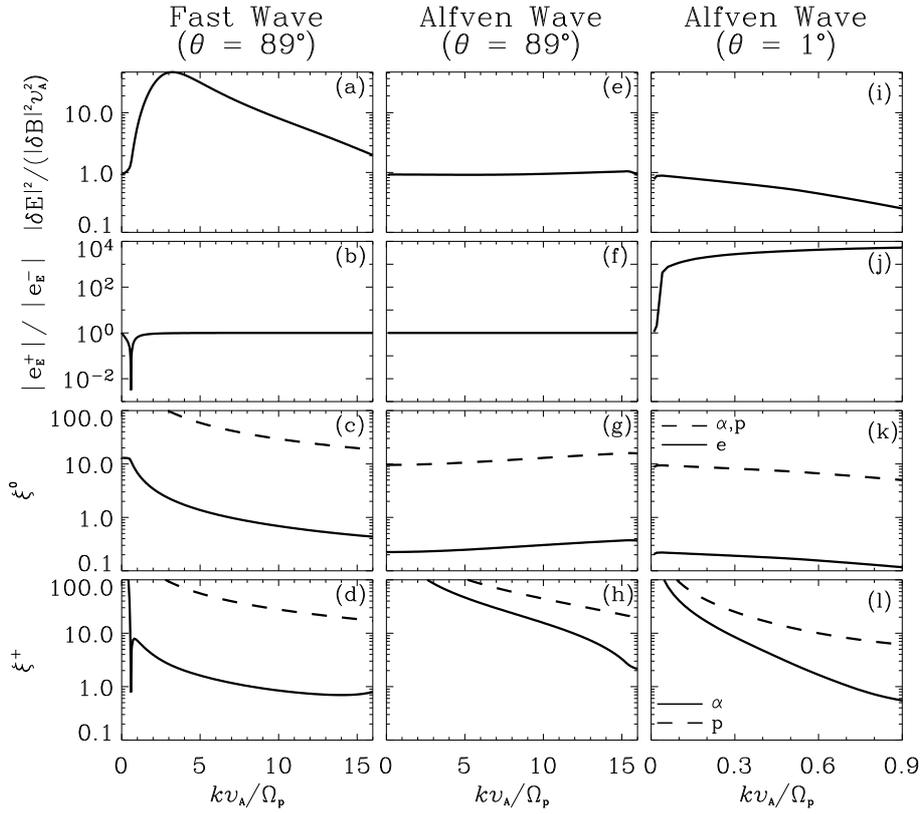}
\caption{Properties of quasi-perpendicular fast waves,
quasi-perpendicular and quasi-parallel Alfv\'{e}n waves in a
proton/alpha/electron plasma: (a,e,i) ratio of electric to
magnetic field turbulence spectra $|\delta \mathbf{E}|^2 /
\,|\delta \mathbf{B}|^2$, (b,f,j) ratio of left- to right-handed
polarizations for electric field spectrum $|\mathbf{e}_E^+| \,/ \,
|\mathbf{e}_E^-|$, (c,g,k) Landau resonance factors $\xi^0$, and
(d,h,l) cyclotron resonance factors $\xi^+$.} \label{EB}
\end{figure}

\begin{figure}
\includegraphics[width=.99\textwidth]{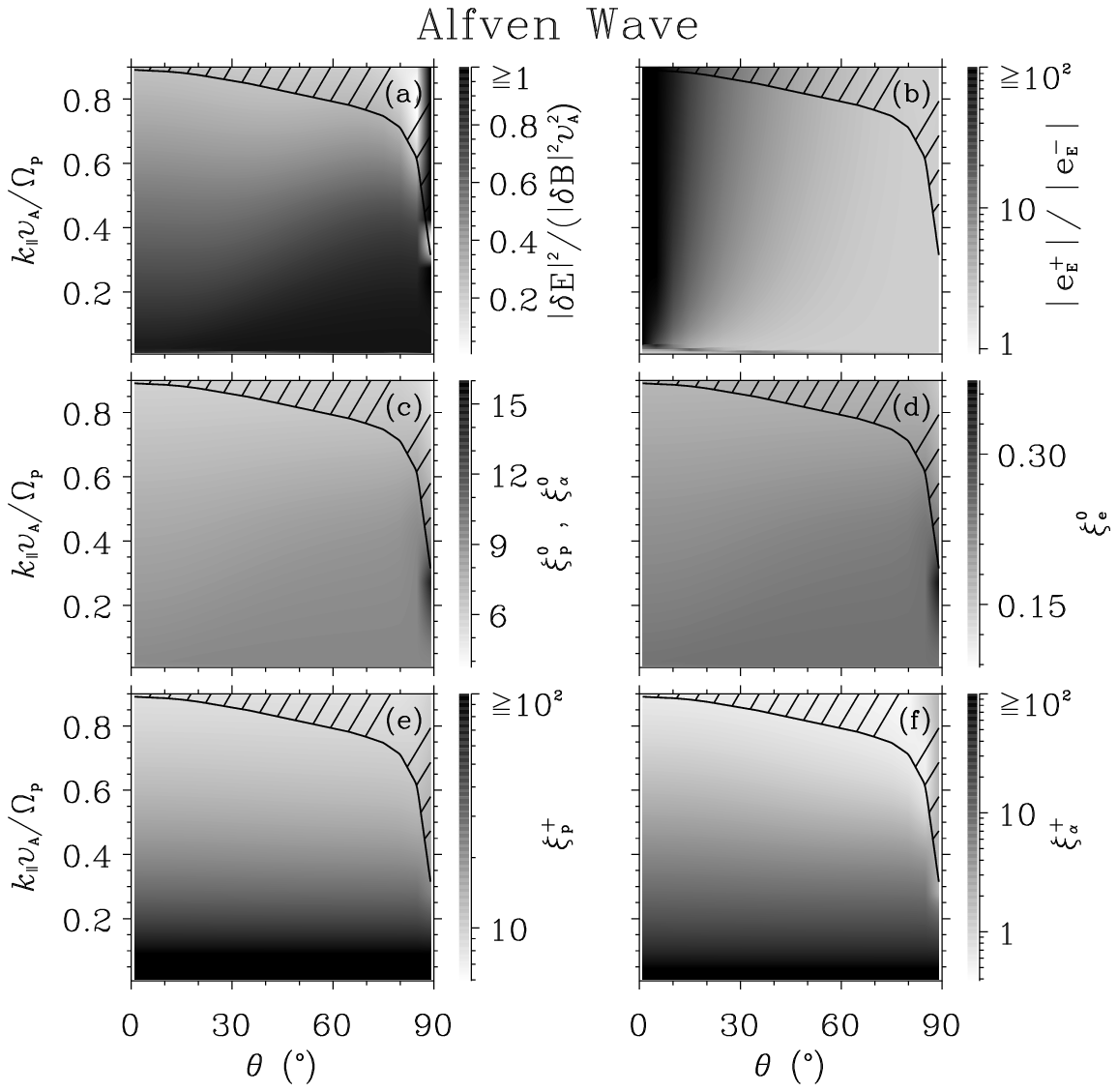}
\caption{The dependence of Alfv\'{e}n wave properties in a
proton/alpha/electron plasma on the parallel wave number
$k_\parallel$ and propagation angle $\theta$: (a) ratio of
electric to magnetic field turbulence spectra $|\delta
\mathbf{E}|^2 / \, |\delta \mathbf{B}|^2$, (b) ratio of left- to
right-handed polarizations for electric field spectrum
$|\mathbf{e}_E^+| \,/ \, |\mathbf{e}_E^-|$, (c) Landau resonance
factors of protons $\xi^0_{\rm p}$ and alpha particles
$\xi^0_\alpha$, (d) Landau resonance factor of electrons
$\xi^0_{\rm e}$, (e) cyclotron resonance factor of protons
$\xi^+_{\rm p}$, and (f) cyclotron resonance factor of alpha
particles $\xi^+_\alpha$.} \label{EB-Alfven}
\end{figure}

\begin{figure}
\includegraphics[width=0.99\textwidth]{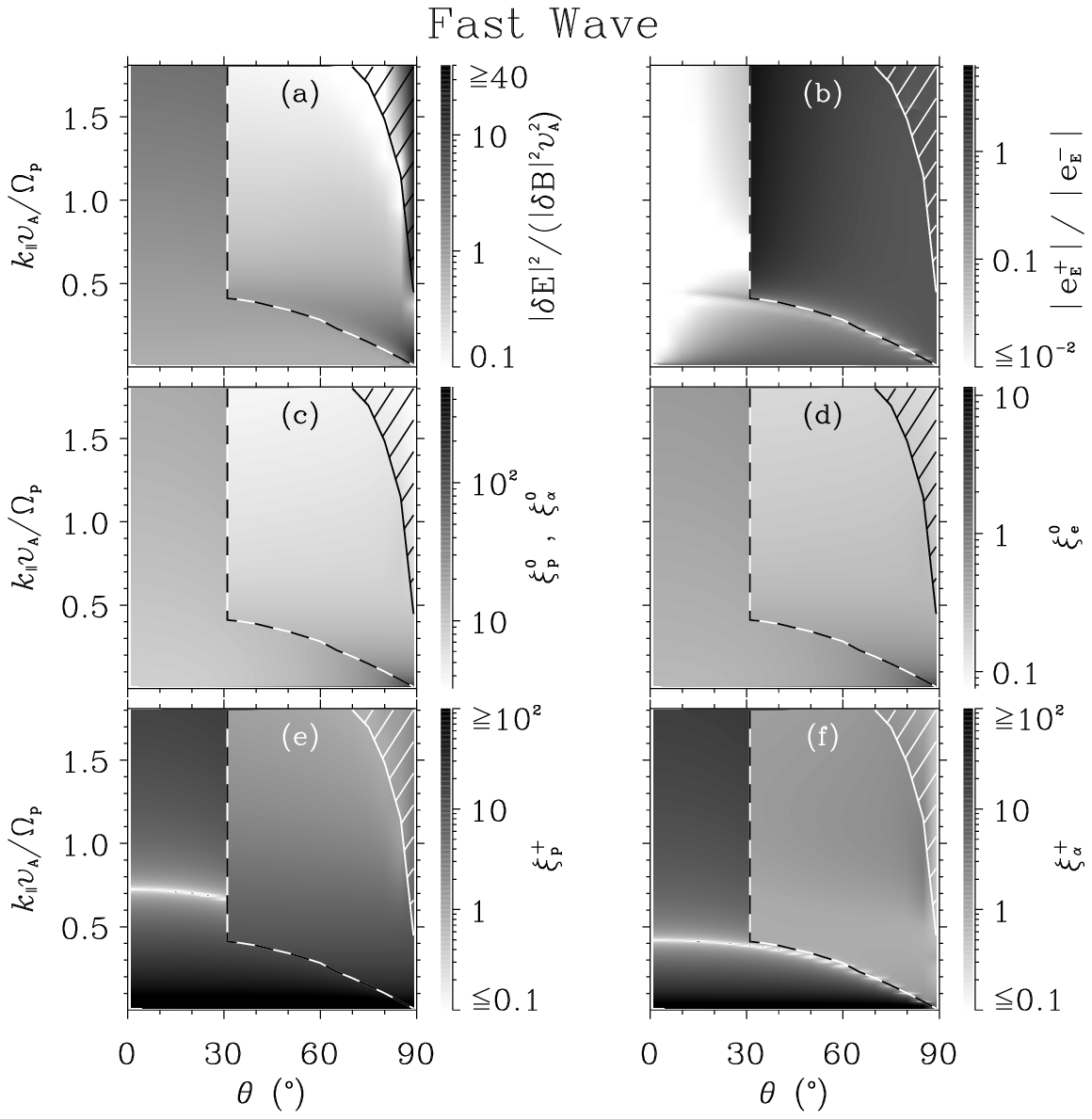}
\caption{The dependence of fast wave parameters in a
proton/alpha/electron plasma on the parallel wave number
$k_\parallel$ and propagation angle $\theta$: (a) ratio of
electric to magnetic field turbulence spectra $|\delta
\mathbf{E}|^2 /\, |\delta \mathbf{B}|^2$, (b) ratio of left- to
right-handed polarizations for electric field spectrum
$|\mathbf{e}_E^+| \,/ \, |\mathbf{e}_E^-|$, (c) Landau resonance
factors of protons $\xi^0_{\rm p}$ and alpha particles
$\xi^0_\alpha$, (d) Landau resonance factor of electrons
$\xi^0_{\rm e}$, (e) cyclotron resonance factor of protons
$\xi^+_{\rm p}$, and (f) cyclotron resonance factor of alpha
particles $\xi^+_\alpha$.} \label{EB-Fast}
\end{figure}

Electromagnetic behaviors of Alfv\'{e}n and fast waves are
displayed in Figures \ref{EB}, \ref{EB-Alfven}, and \ref{EB-Fast}.
As fast-cyclotron and Alfv\'{e}n-cyclotron waves become more
oblique, phase speeds $\omega_{\rm r} / k$ involving cyclotron
resonances are gradually reduced towards 0, and electric field
oscillations $\delta \mathbf{E}$ become more aligned with wave
vector $\mathbf{k}$. The values of $\omega_{\rm r} / k$ and ($1 -
|\mathbf{e}_k \cdot \mathbf{e}_E|$) determine the ratio of
electric to magnetic field fluctuations $|\delta \mathbf{E}|^2 /
\, |\delta \mathbf{B}|^2$ (Equation (\ref{Equ:mag2})). For
Alfv\'{e}n-cyclotron waves with $0^\circ \le \theta < 90^\circ$
(Figures \ref{EB}a and \ref{EB-Alfven}a) and fast-cyclotron waves
with $31^\circ \le \theta < 90^\circ$ (Figures \ref{EB}e and
\ref{EB-Fast}a), $|\delta \mathbf{E}|^2 /\, |\delta \mathbf{B}|^2$
increases when $\theta$ becomes larger. At oblique enough angles,
both fast and Alfv\'{e}n waves are nearly electrostatic and
linearly polarized as a result of $\delta \mathbf{E}$ being almost
parallel to $\mathbf{k}$. In particular, nearly-perpendicular fast
waves become almost electrostatic, as $|\delta \mathbf{E}|^2 /
(|\delta \mathbf{B}|^2 \, v^2_{\rm A})$ could be as high as 50 at
$k\, v_{\rm A} / \,\Omega_{\rm p} =3$ (Figures \ref{EB}a and
\ref{EB-Fast}a). As a result, electric field fluctuations $|\delta
\mathbf{E}|$ of fast waves are significantly enhanced over their
corresponding magnetic field fluctuations $|\delta \mathbf{B}|$.
As for Alfv\'{e}n waves, $|\delta \mathbf{E}|$ and $|\delta
\mathbf{B}|\cdot v_{\rm A}$ are equally strong for
quasi-perpendicular propagation (Figures \ref{EB}e and
\ref{EB-Alfven}a), though $|\delta \mathbf{E}|$ is far less than
$|\delta \mathbf{B}|\cdot v_{\rm A}$ at large $k$ for parallel
propagation (Figures \ref{EB}i and \ref{EB-Alfven}a). Meanwhile,
for oblique propagation, parallel electric field fluctuations
$|\delta E_\parallel|$ are enhanced over the whole spectrum. The
overall sense of the electric field polarization is left-handed
for parallel Alfv\'{e}n waves (Figures \ref{EB}j and
\ref{EB-Alfven}b), right-handed for parallel fast waves (Figure
\ref{EB-Fast}b), and nearly linearly polarized for perpendicular
Alfv\'{e}n and fast waves (Figures \ref{EB}b, \ref{EB}f,
\ref{EB-Alfven}b, and \ref{EB-Fast}b). Obviously, linear mode
conversion from fast waves to alpha cyclotron waves near alpha
cyclotron frequency $\Omega_\alpha$ abruptly increases the
left-handed electric field oscillation (Figure \ref{EB-Fast}b).
When present in the inner corona, both Alfv\'{e}n and fast waves
could heat plasma species through Landau and cyclotron resonances.
If a resonant factor $|\xi_j|$ is large, only particles in the
remote tail of the VDF of species $j$ are in resonance with the
wave. In linear Vlasov theory, a resonance factor of $|\xi_j| \leq
3$ is a necessary but not a sufficient condition for
non-negligible wave-particle interaction \cite{Gary2004b}. Both
$|\xi_j^0|\leq 3$ and parallel electric field oscillation $|\delta
E_\parallel|>0$ are required for noticeable Landau resonance;
Similarly, both $|\xi_j^+| \le 3$ and left-handed electric field
oscillation $|\delta E^+| >0$ correspond to cyclotron resonance.
Due to nearly zero $|\delta E_\parallel|$ accompanying
quasi-parallel waves, Landau resonances are very weak even with
the synchronous condition of wave and particle phases, {\it i.e.}
$|\xi^0_j| \le 3$. Conspicuous streaks of $\xi^+_{\rm p}$ from
$0^\circ < \theta < 31^\circ$ (Figure \ref{EB-Fast}e) and
$\xi^+_\alpha$ from $0^\circ < \theta < 90^\circ$ (Figure
\ref{EB-Fast}f) correspond to locally enhanced damping of fast
waves at ion cyclotron frequencies $\Omega_\alpha$ and
$\Omega_{\rm p}$. However, as explained by \inlinecite{li2001},
the localized cyclotron damping of fast waves by species $j$ is
unlikely to be important because of its localized occurrence in a
very narrow frequency band across $\Omega_j$. Through Landau
and/or cyclotron resonances, different plasma species absorbs
energy from the turbulence power spectrum $|\delta
\mathbf{B}(\mathbf{k})|^2$ accordingly.

\section{Effect of Wave Propagation Angle}\label{Sec:angle}
\begin{figure}
\includegraphics[height=.99\textwidth,angle=90]{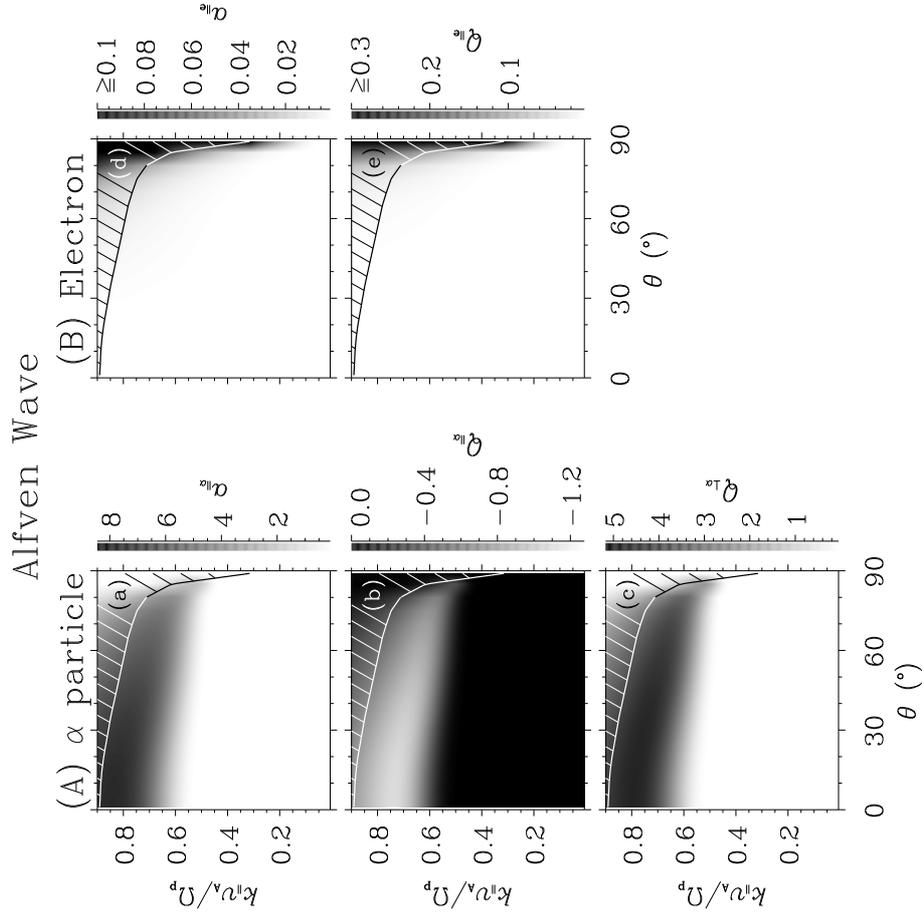}
\caption{The energization capabilities of Alfv\'{e}n waves in a
proton/alpha/electron plasma on (A) alpha particles and (B)
electrons. From top to bottom are (a,d) parallel acceleration
$a_\parallel$, (b,e) parallel heating $Q_{\parallel}$, and (c)
perpendicular heating $Q_{\perp}$ over the $k_\parallel - \theta$
domain. Quantities of $a_\parallel$, $Q_{\parallel}$, and
$Q_{\perp}$ are already normalized by the wave spectrum $|\delta
\mathbf{B}(\mathbf{k})|^2$. Here the proton energization and the
perpendicular electron heating are essentially
zero.}\label{aQ-tht-Alfven}
\end{figure}

\begin{figure}
\includegraphics[height=.99\textwidth,angle=90]{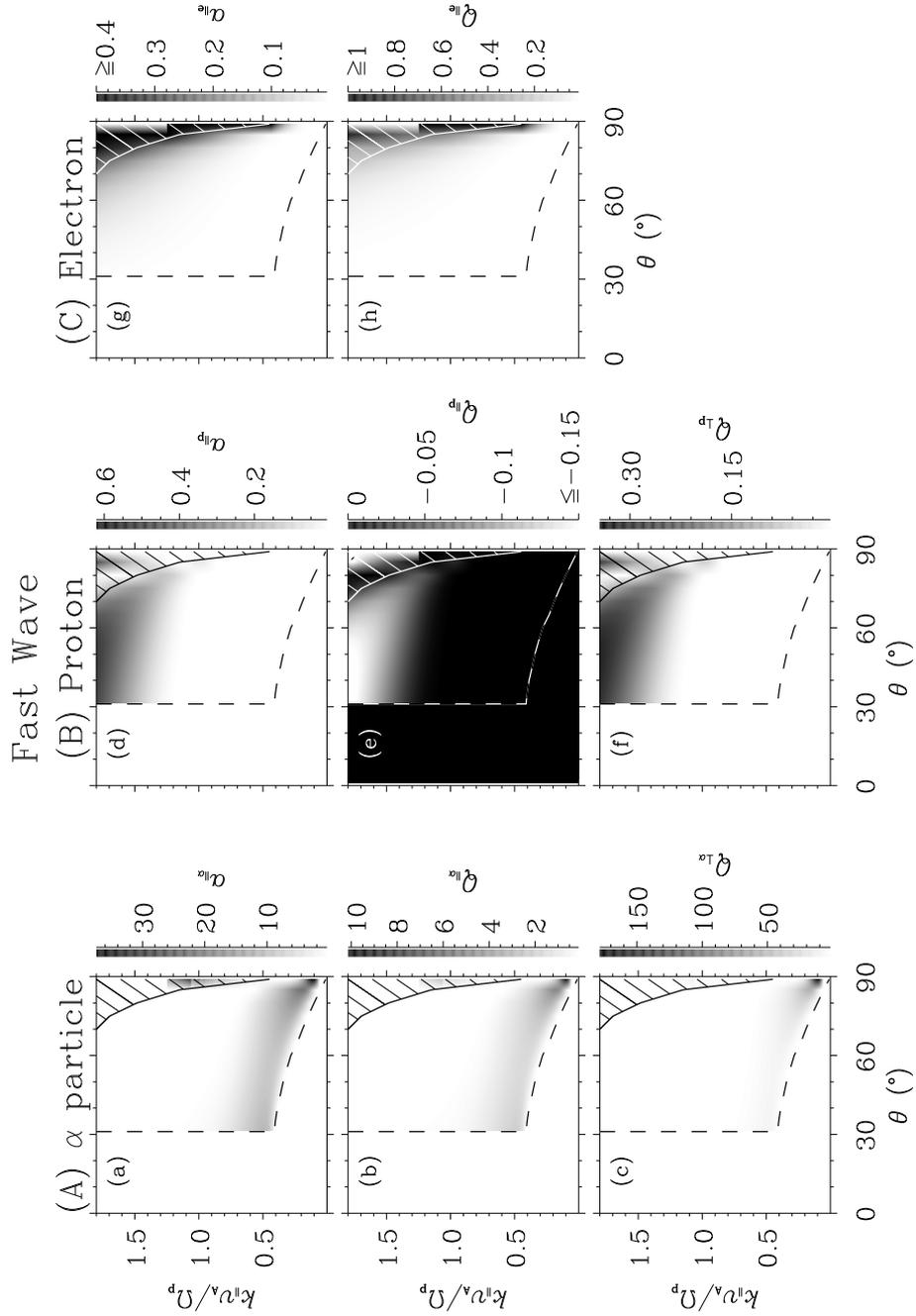}
\caption{The energization effects of fast waves in a
proton/alpha/electron plasma on (A) alpha particles, (B) protons,
(C) electrons. From top to bottom are (a,d,g) parallel
acceleration, (b,e,h) parallel heating, and (c,f) perpendicular
heating over the $k_\parallel - \theta$ domain. Here the
perpendicular electron heating $Q_{\perp {\rm e}}$ is essentially
zero.}\label{aQ-tht-Fast}
\end{figure}

The energization of any plasma species $j$ due to turbulence PSD
$|\delta \mathbf{B}(\mathbf{k})|^2$ is estimated by quasi-linear
transport coefficients ($a_{\parallel j}$, $Q_{\parallel j}$, and
$Q_{\perp j}$) in Equation (\ref{Equ:aQ}) and visualized in
Figures \ref{aQ-tht-Alfven} and \ref{aQ-tht-Fast}. Alfv\'{e}n and
fast waves at each $\mathbf{k}$ are assumed to be equally strong;
the magnitude of fluctuation $|\delta \mathbf{B}|$ is the same for
both waves. The parallel acceleration $a_{\parallel j}$, parallel
heating $Q_{\parallel j}$, and perpendicular heating $Q_{\perp j}$
are basically determined by the resonance function $R$ in Equation
(\ref{Equ:Rjkn}). Different species show different behaviors in
terms of their energization from wave damping.
Alfv\'{e}n-cyclotron waves essentially do not interact with
protons, because their asymptotic frequency $\Omega_\alpha$ is far
away from the proton cyclotron frequency $\Omega_{\rm p}$ (Figure
\ref{wk-Alfven}a). Fast-whistler waves with propagation angles
$0^\circ \le \theta < 31^\circ$ are essentially undamped, as shown
in Figure \ref{aQ-tht-Fast}. Noticeable energy recipients of
Alfv\'{e}n waves are alpha particles when $0^\circ \le \theta <
80^\circ$ and electrons when $80^\circ \le \theta < 90^\circ$
(Figure \ref{aQ-tht-Alfven}); those of fast waves are alpha
particles with $31^\circ \le \theta < 90^\circ$, protons with
$31^\circ \le \theta < 80^\circ$, and electrons with $80^\circ \le
\theta < 90^\circ$ (Figure \ref{aQ-tht-Fast}). The alpha cyclotron
waves in Figure \ref{aQ-tht-Alfven} have similar energization
effects as the proton cyclotron waves in Figure \ref{aQ-tht-Fast},
though being damped more heavily. The strongest energy recipient
among all plasma species is the alpha particles {\it via} alpha
cyclotron resonances with fast-cyclotron waves when $31^\circ \le
\theta < 90^\circ$ (Figures \ref{aQ-tht-Fast}a-c). Such an alpha
energization from fast-cyclotron waves is increasingly
significant, as the waves become more oblique towards
$\theta=90^\circ$. Moreover, for effective energization from
fast-cyclotron waves, alpha particles occupy a lower part of the
$k_\parallel - \theta$ parameter space than either protons or
electrons. Based on the turbulence cascade scenario, the PSD is
absorbed by resonant alpha particles from fast-cyclotron waves at
lower frequencies and wave numbers, and hence has larger
fluctuation amplitudes $|\delta \mathbf{B}|^2$. As a result, the
combination of the largest $|\delta \mathbf{B}|^2$ in Equation
(\ref{Equ:Marsch}) and the largest ($a_{\parallel \alpha}$,
$Q_{\parallel \alpha}$, $Q_{\perp \alpha}$) in Equation
(\ref{Equ:aQ}) are responsible for the strongest alpha
energization. In addition, compared with alpha cyclotron frequency
$\Omega_\alpha$, frequencies $\omega_{\rm r}$ at the alpha
cyclotron resonance are slightly higher for fast-cyclotron waves
(Figure \ref{wk}c) and slightly lower for Alfv\'{e}n-cyclotron
waves (Figures \ref{wk}a and \ref{wk}c). Consequently, the
first-order parallel resonant speeds $w_\alpha(\mathbf{k}, 1)$ are
positive for fast-cyclotron waves and negative for
Alfv\'{e}n-cyclotron waves (Equation (\ref{Equ:Doppler})).
According to Equations (\ref{Equ:Marsch}) and (\ref{Equ:aQ}),
fast-cyclotron and Alfv\'{e}n-cyclotron waves are responsible for
parallel heating and cooling of alpha particles respectively. With
Alfv\'{e}n waves, it is much easier for the VDF of alpha particles
to develop into a kinetic anisotropy of $T_{\perp \alpha} >
T_{\parallel \alpha}$, and then trigger an ion cyclotron
instability of alpha particles. In a cold plasma of $\beta_{\rm
p}=0.01$, alpha particles are the primary energy recipient from
the damping of both Alfv\'{e}n and fast waves. This preferential
energization of alpha particles is much more significant by
oblique fast-cyclotron waves {\it via} linear mode conversion from
their corresponding fast-whistler waves.

\section{Influence of Minor Ion Presence}\label{Sec:minor-ion}
Without alpha particles in a proton/electron plasma, oblique
fast-whistler waves can still be linearly converted to Bernstein
waves at harmonic proton cyclotron frequencies of $\Omega_{\rm p}$
and $2 \, \Omega_{\rm p}$ \cite{li2001,Markovskii2010}. At a
perpendicular angle of $\theta=90^\circ$, Bernstein waves are
purely electrostatic, do not damp, and cannot affect ions
\cite{Stix1992}. When $\theta$ deviates from $90^\circ$, Bernstein
waves are strongly damped. With linear mode coupling, highly
oblique fast-Bernstein waves are strongly dissipated by protons,
as previously demonstrated from linear Vlasov analyses
\cite{li2001} and hybrid simulation \cite{Markovskii2010} of
fast-wave cascade spectrum. With the absence of alpha particles,
energization rates of protons and electrons are estimated by
Equation (\ref{Equ:aQ}). Only differing in the abundance of alpha
particles $n_\alpha = 0.02 \, n_{\rm p}$ (Tables \ref{Tab2} and
\ref{Tab3}), the two sets of initial plasma parameters in Sections
\ref{Sec:angle} and \ref{Sec:minor-ion} result in dramatically
different energizations. Linear mode conversion from fast to
Bernstein waves occurs, when the fast waves are oblique enough,
{\it i.e.}, propagation angles $\theta \ge 64^\circ$.
Specifically, such a conversion point of fast-Bernstein wave
branch is at $2 \, \Omega_{\rm p}$ for $64^\circ \leq \theta <
87^\circ$ and $\Omega_{\rm p}$ for $87^\circ \leq \theta <
90^\circ$. Energization of protons and electrons are noticeable
for $64^\circ \leq \theta < 85^\circ$ and $80^\circ \leq \theta <
90^\circ$, respectively. Because of the linear mode coupling,
fast-whistler waves are intercepted at frequencies $\omega_{\rm i}
\approx \Omega_\alpha$ for alpha cyclotron waves and $\omega_{\rm
i} \approx \Omega_{\rm p}$ (or $2 \, \Omega_{\rm p}$) for
Bernstein waves. Parallel electric field $|\delta E_\parallel|$
between $\Omega_\alpha$ and $\Omega_{\rm p}$ of fast-Bernstein
waves are responsible for a much stronger electron energization in
contrast with fast-cyclotron waves. Protons undergo larger
energization by fast-Bernstein waves than by fast-cyclotron waves,
particularly in perpendicular heating $Q_{\perp {\rm p}}$.
However, fast-cyclotron waves primarily energize alpha particles,
and such a strong ion energization cannot be found in
fast-Bernstein waves. When alpha particles are present, fast waves
are coupled to cyclotron waves instead of Bernstein waves as a
result of the turbulence cascade. Approaching alpha cyclotron
frequency $\Omega_\alpha$ (instead of $\Omega_{\rm p}$), a
fast-cyclotron wave has a smaller frequency $\omega_{\rm r}$, a
smaller wave number $k$, and larger wave energy density $|\delta
\mathbf{B}(\mathbf{k})|^2$ based on the turbulence cascade
scenario. Moreover, the angle threshold for linear mode coupling
is low for fast-cyclotron waves ($\theta \ge 31^\circ$) and high
for fast-Bernstein waves ($\theta \ge 64^\circ$). Therefore, the
existence of alpha particles could lead to more efficient damping
of fast waves.

\section{Role of Relative Flow of Minor Ions} \label{Sec:dif-V}
\begin{figure}
\includegraphics[height=.99\textwidth,angle=90]{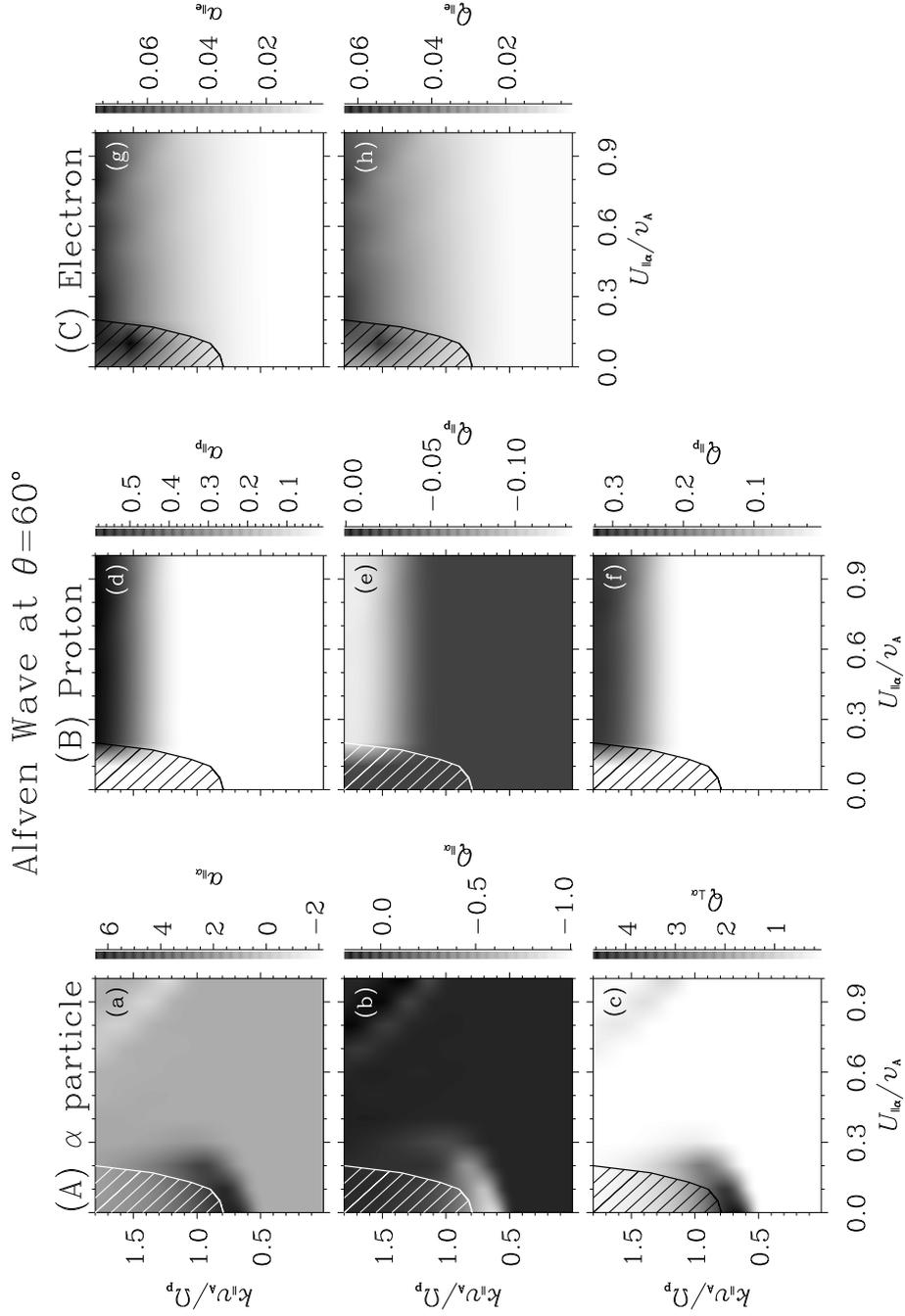}
\caption{For Alfv\'{e}n waves at propagation angle $\theta =
60^\circ$, differential speed of alpha particles $U_{\parallel
\alpha}$ in a proton/alpha/electron plasma results in different
energizations of (A) alpha particles, (B) protons, and (C)
electrons.}\label{aQ-V-Alfven}
\end{figure}

\begin{figure}
\includegraphics[height=.99\textwidth,angle=90]{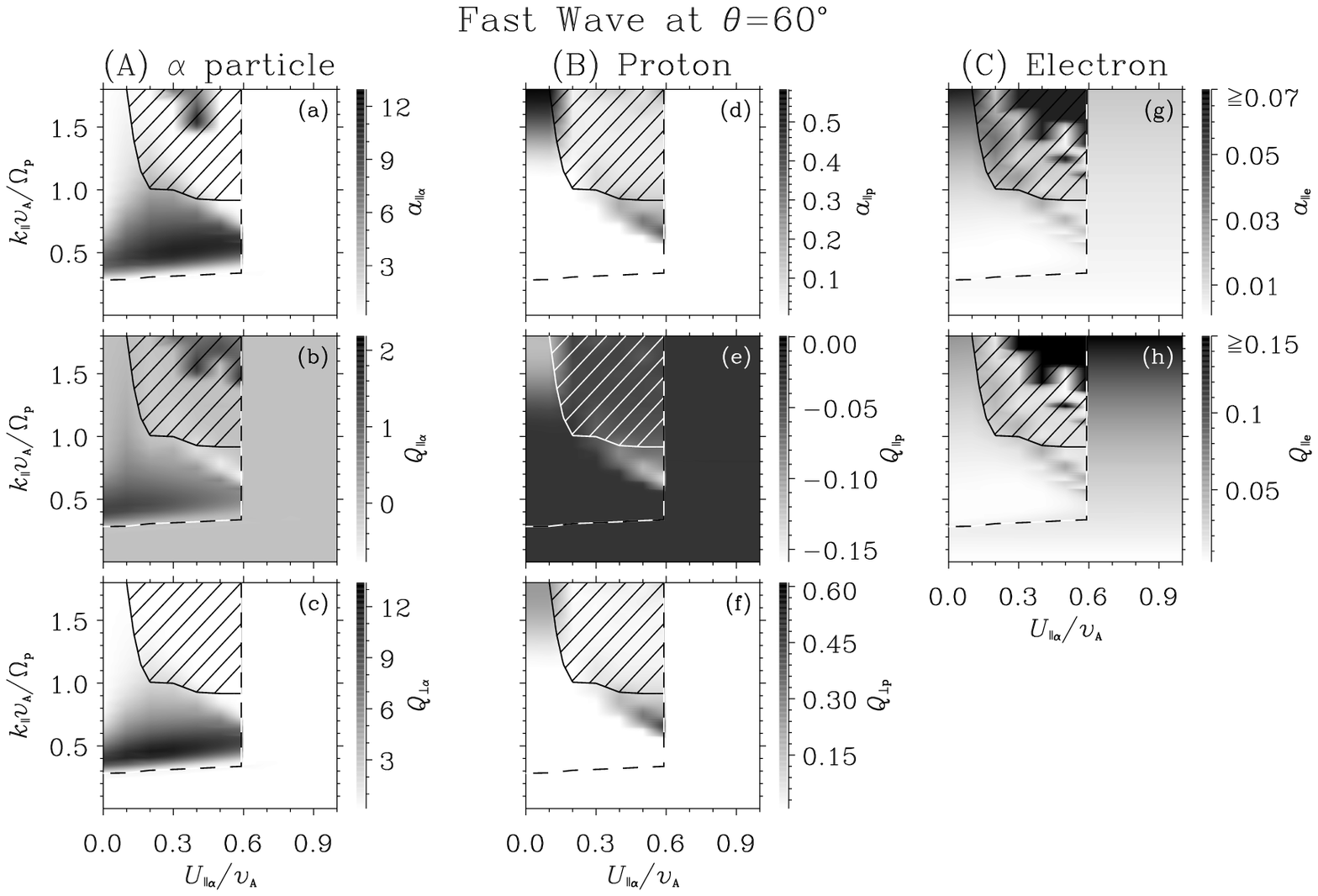}
\caption{Depending on the differential speed of alpha particles
$U_{\parallel \alpha}$ in a proton/alpha/electron plasma, the
energization consequence of fast waves at propagation angle
$\theta = 60^\circ$. Linear mode conversion occurs at
$U_{\parallel \alpha} < 0.6 \, v_{\rm A}$, denoted as a
black-and-white line in the $k_\parallel - U_{\parallel \alpha}$
domain.}\label{aQ-V-Fast}
\end{figure}

Wave-particle interaction in a multi-ion plasma may be
significantly modulated by a differential speed $U_{\parallel j}$
of minor ion $j$. Here $U_{\parallel j}$ is defined as the bulk
flow speed relative to protons. The dependence of energization on
$U_{\parallel \alpha}$ in a proton/alpha/electron plasma is
presented in Figure \ref{aQ-V-Alfven} for Alfv\'{e}n waves and
Figure \ref{aQ-V-Fast} for fast waves along propagation angle
$\theta=60^\circ$. Initially cold alpha particles are swiftly
accelerated and heated by both Alfv\'{e}n-cyclotron and
fast-cyclotron waves. Fast-cyclotron waves resonate with alpha
particles at frequencies $\omega_{\rm i} \approx \Omega_\alpha$
and protons at $\omega_{\rm i} \approx \Omega_{\rm p}$, and
energize alpha particles at a low wave number $k$ (Figures
\ref{aQ-V-Fast}a-c) and protons at a high $k$ (Figures
\ref{aQ-V-Fast}d-f). Obviously, alpha particles are energized much
more strongly than protons for Alfv\'{e}n-cyclotron waves at
$U_{\parallel \alpha} < 0.2 \, v_{\rm A}$ (Figure
\ref{aQ-V-Alfven}) and for fast-cyclotron waves at $U_{\parallel
\alpha} < 0.6 \, v_{\rm A}$ (Figure \ref{aQ-V-Fast}). When
$U_{\parallel \alpha}$ reaches 0.2 $v_{\rm A}$,
Alfv\'{e}n-cyclotron waves at $\theta=60^\circ$ begin to resonate
with protons, and alpha particles lose the initial energization
advantage. Meanwhile, oblique fast-cyclotron waves continue to
preferentially energize alpha particles. The appearance of
fast-cyclotron waves occurs for $\theta \ge 31^\circ$ at
$U_{\parallel \alpha}/ \,v_{\rm A}=0$ (Figure \ref{aQ-tht-Fast})
and $\theta > 60^\circ$ at $U_{\parallel \alpha}/ \,v_{\rm A}=0.6$
(Figure \ref{aQ-V-Fast}). Obviously, as a response to continuously
increasing $U_{\parallel \alpha}$, the minimum threshold of wave
propagation angle $\theta$ at which the coupling between
fast-whistler and hybrid alpha-proton cyclotron waves occurs is
increased accordingly. As a result of continuous preferential
energization, alpha particles gradually accumulate high enough
$U_{\parallel \alpha}$ and temperature anisotropy $T_{\perp
\alpha}/\, T_{\parallel \alpha}$. When alpha particles lose
resonance, the major wave-particle interaction is proton cyclotron
resonance for Alfv\'{e}n-cyclotron waves (Figures
\ref{aQ-V-Alfven}d-f) and electron Landau resonance for
fast-whistler waves (Figure \ref{aQ-V-Fast}h). Productions of much
higher $U_{\parallel \alpha}$ and $T_{\perp \alpha}/\,
T_{\parallel \alpha}$ are inhibited, as excessive free energy of
alpha beam and kinetic anisotropy in the VDF of alpha particles
would be released to excite Alfv\'{e}n-cyclotron waves (Figures
\ref{aQ-V-Alfven}a-c). The self-consistent evolution of alpha
particles in the parametric space of $U_{\parallel \alpha}$ and
$T_{\perp \alpha}/\, T_{\parallel \alpha}$ could be further
investigated by particle simulations. Here from Vlasov analyses of
various plasma parameters, both fast and Alfv\'{e}n waves are
likely to contribute to the eventually sustainable differential
speed and temperature anisotropy of minor ions in the solar wind.

\section{Dependence on Plasma Temperature} \label{Sec:temperature}
Fast-cyclotron waves only exist in a cold plasma, because linear
coupling of fast-whistler and hybrid-cyclotron waves disappears
for large plasma beta. Parametric studies of proton beta
$\beta_{\rm p}$ are performed for Alfv\'{e}n and fast waves along
propagation angle $\theta=60^\circ$ (Tables \ref{Tab2} and
\ref{Tab3}). Generally, in an increasingly hot plasma, waves are
increasingly damped, and their effective $\omega_{\rm r} - k$
dispersion curves are consequently shortened according to Equation
(\ref{Equ:Gary-threshold}). $\mbox{Max.}(k_\parallel)$ generally
decreases with an increasing $\beta_{\rm p}$, which is estimated
at each $\beta_{\rm p}$ in the $k_\parallel - \beta_{\rm p}$
parameter domain according to weak damping condition (Equation
(\ref{Equ:Gary-threshold})). Exceptional increases of
$\mbox{Max.}(k_\parallel)$ are found from a transition of $-1.2 <
\log_{10}(\beta_{\rm p}) < -1$ and a jump across
$\log_{10}(\beta_{\rm p})= -0.95$. Such exceptions in the trend of
$\mbox{Max.}(k_\parallel)$ versus $\beta_{\rm p}$ are due to a
significant dispersion change of corresponding wave branch. As
$\beta_{\rm p}$ increases, the dispersion of Alfv\'{e}n waves
finally changes to cross the alpha cyclotron frequency
$\Omega_\alpha$, and fast waves ultimately cease the conversion to
hybrid alpha-proton waves. The sudden change of wave dispersion
corresponds to a certain value of $\beta_{\rm p}$. The critical
point is $\log_{10}(\beta_{\rm p})= -1.2$ for Alfv\'{e}n waves and
$\log_{10}(\beta_{\rm p})= -0.95$ for fast waves. Across the
critical points, wave branches extend to higher frequencies
$\omega_{\rm r}$, and survive for higher wave numbers
$k_\parallel$. Alpha particles are the chief energy recipient from
both Alfv\'{e}n-cyclotron waves and fast-cyclotron waves ($\,
\log_{10}(\beta_{\rm p}) < -0.95$). The energization effects of
Alfv\'{e}n-cyclotron waves weaken as $\beta_{\rm p}$ increases,
while those of fast-cyclotron waves peak in the range of $-2.2 <
\log_{10}(\beta_{\rm p}) < -1.2$. In addition, without mode
conversion for $\log_{10}(\beta_{\rm p}) \ge -0.95$, oblique
fast-whistler waves not only energize ions around ion cyclotron
frequencies ($\Omega_\alpha$ and $\Omega_{\rm p}$), but also
dissipate a noticeable portion of wave energy to electrons in the
form of parallel heating $Q_{\parallel {\rm e}}$. To the energy
absorption of fast-whistler waves, the contribution of electrons
is comparable to that of protons and alpha particles in a warm
plasma. Only in a cold plasma, linear coupling of fast and
cyclotron waves exists, and the electron heating could be
negligible for damping of fast-cyclotron waves. Hence, in the
inner corona with $\beta_{\rm p} \approx 0.01$, oblique
fast-cyclotron waves are expected to play a more important role in
energizing minor ions, provided that they are equally as strong as
Alfv\'{e}n-cyclotron waves.

\section{Discussion on Nonlinear Stage of Wave-Particle Interaction} \label{Sec:PSD}
It is still an enigma how the macro-scale heating and acceleration
of the solar wind are powered by waves on ion kinetic scales and
how these waves are generated by multi-scale coupling. The solar
wind turbulence becomes fully developed on a time scale that is
very short, compared to the expansion time of the solar wind along
open magnetic flux tubes. So the turbulence evolution in the wave
vector space could be considered to develop in a spatially
homogeneous plasma, so we can adopt spatially uniform energization
rates (Equation (\ref{Equ:aQ})) in this paper (see also
\inlinecite{Cranmer2003}). In Equation (\ref{Equ:aQ}),
$a_{\parallel j}$ is probably not an important contributor to the
solar wind acceleration, as the magnetic mirror force due to
anisotropic heating of $Q_{\parallel j}$ and $Q_{\perp j}$ is
typically stronger. A plasma species $j$ is energized by
dissipation of its ion cyclotron waves. In a frame moving with the
phase speed of the waves, a diffusion in the VDF of species $j$
(Equation (\ref{Equ:VDF})) occurs along contours of constant
energy. Such a diffusion process is faster for species $j$ with a
lower charge-to-mass ratio. This wave-particle interaction was
formulated into a kinetic shell model \cite{Isenberg2001}, which
quantitatively describes ion energization by bidirectionally
propagating ion cyclotron waves. As a result, the VDF in Equation
(\ref{Equ:VDF}) is gradually deviated from its initially
Maxwellian distribution. The quasi-linear coefficients of Equation
(\ref{Equ:Marsch}) are only valid in a small deviation from the
Maxwellian VDF of particles (Equation (\ref{Equ:VDF})) and a small
damping rate of plasma waves (Equation
(\ref{Equ:Gary-threshold})). Owing to these limitations, the
analytical method in this paper can only describe the initial
quasi-linear stage of wave-particle interaction. The ensuing
nonlinear stage is beyond the scope of this paper, though it can
be self-consistently described by full particle simulation.

\section{Conclusions and Summary}\label{Sec:Conclusion}
Using the linear Vlasov wave theory and quasi-linear resonant
wave-particle interaction for a low-beta multi-species plasma
\cite{li2001,Marsch2001}, fast and Alfv\'{e}n waves (and their
dispersion relations and electromagnetic field fluctuations) are
analyzed, and then applied to energize a proton/electron/alpha
plasma with a low proton beta of $\beta_{\rm p} \approx 0.01$. In
this paper, we assume that (1) low-frequency Alfv\'{e}n and fast
waves are launched from the solar surface with the same spectral
shape and the same amplitude of PSD, (2) these macro-scale
waves/turbulences continuously cascade towards kinetic scales at
ion cyclotron frequencies and ion inertial scales, and (3) the
transported kinetic-scale waves energize solar wind species
through cyclotron and Landau resonances. If the wave propagation
angle $\theta$ is oblique with respect to the background magnetic
field, fast-whistler waves can be linearly coupled with hybrid
alpha-proton cyclotron waves. Both fast-cyclotron and
Alfv\'{e}n-cyclotron waves are important for preferential heating
and acceleration of minor ions. In this paper, noticeable
energization of alpha particles requires oblique propagation for
fast-cyclotron waves ($\theta \ge 31^\circ$) and non-perpendicular
propagation for Alfv\'{e}n-cyclotron waves ($\theta < 80^\circ$).
Nearly perpendicular Alfv\'{e}n-cyclotron waves at $80^\circ \leq
\theta < 90^\circ$ significantly degenerate their capabilities of
energizing minor ions due to very large perpendicular wave numbers
at ion cyclotron frequencies. Only through complex nonlinear
processes, nearly perpendicular Alfv\'{e}n waves may heat ions.
For instance, \inlinecite{Cranmer2003} proposed that the damping
of kinetic Alfv\'{e}n waves successively leads to electron beams,
Langmuir turbulence, and Debye-scale electron phase space holes,
and ions are perpendicularly heated {\it via} a collision-like
scenario. Meanwhile, oblique fast-cyclotron waves have a larger
left-handedly polarized electric field $|\delta \mathbf{E}^+|$
than the corresponding Alfv\'{e}n-cyclotron waves, because the
total electric field fluctuation $|\delta \mathbf{E}|$ of fast
waves is stronger enough to over-compensate its smaller ratio of
left- to right-handedly polarized electric fields $|\delta
\mathbf{E}^+| / |\delta \mathbf{E}^-|$. Moreover, in contrast to
Alfv\'{e}n-cyclotron waves, alpha cyclotron resonance happens at a
smaller wave number for fast-cyclotron waves especially at $\theta
\approx 90^\circ$, where the turbulence PSD has a larger local
amplitude. Hence, fast-cyclotron waves have a greater capability
of energizing solar wind ions than Alfv\'{e}n-cyclotron waves.
Interacting with Alfv\'{e}n-cyclotron and fast-cyclotron waves,
alpha particles are preferentially energized as the primary
recipient of wave energy. In contrast to alpha particles,
electrons are generally less energized. When the differential
speed $U_{\parallel \alpha}$ of alpha particles is high enough,
Alfv\'{e}n-cyclotron waves lose alpha cyclotron resonance much
earlier than fast-cyclotron waves. For example, at $\theta =
60^\circ$, alpha cyclotron resonance disappears at $U_{\parallel
\alpha} = 0.2 \, v_{\rm A}$ for Alfv\'{e}n-cyclotron waves and
$U_{\parallel \alpha} = 0.6 \, v_{\rm A}$ for fast-cyclotron
waves. When alpha particles lose cyclotron resonance,
fast-cyclotron waves restore to fast-whistler waves without linear
mode conversion, and Alfv\'{e}n-cyclotron waves begin to mainly
energize protons. Moreover, an extremely fast beam and/or a high
temperature anisotropy of minor ions could release their excessive
kinetic energy to excite ion cyclotron waves. Therefore both fast
and Alfv\'{e}n waves are likely to contribute to the eventually
sustainable and balanced differential speed and temperature
anisotropy of minor ions in the solar wind.

Linear mode conversion between fast-whistler and hybrid cyclotron
waves due to the presence of minor ions could be an efficient
mechanism to dissipate wave energy into minor ions. In a warm
plasma in the interplanetary space, such mode coupling vanishes,
and parallel electron heating by Landau damping of oblique
fast-whistler waves becomes non-negligible. As demonstrated in
this paper and the previous work of \inlinecite{li2001},
fast-cyclotron waves besides Alfv\'{e}n-cyclotron waves could also
be considered as a promising candidate for preferential heating
and acceleration of minor ions in the inner corona. Since fast and
Alfv\'{e}n waves heat the solar wind ions in very different ways,
it is important to know the exact nature of plasma turbulence at
kinetic scales. In this regard, observation data such as {\it
Cluster} archive can continue to address the issue.

\begin{acks}
This research was supported by a grant from the Science \&
Technology Facilities Council (STFC) to the Aberystwyth
University, UK. We are very grateful to the anonymous referee for
his/her thoughtful and constructive comments, which have greatly
improved the quality of this paper. We also sincerely thank Prof.
Takashi Sakurai and Mr. Jeff Smith for carefully polishing both
language and style of this paper.
\end{acks}


\newpage

\begin{table}
\caption{Symbols used in this paper and their
denotations}\label{Tab1}
\begin{tabular}{ll}
\hline Symbol & Description \\ \hline %
$c$ & light speed \\ \hline %
$e$ & an electron charge \\ \hline %
$k_{\rm B}$ & Boltzmann constant \\ \hline %
$x$, $y$, $z$ & Cartesian coordinate system \\ \hline %
$\mathbf{B_0}$, $B_0$ & uniform background magnetic field along $z$-axis ($B_0 = |\, \mathbf{B_0}|$) \\ \hline %
$\delta \mathbf{B}$ & magnetic field turbulence \\ \hline %
$\delta \mathbf{E}$ & electric field turbulence \\ \hline %
$\parallel$ & parallel direction ($z$-axis) \\ \hline %
$\perp$ & perpendicular direction ($xy$-plane) \\ \hline %
$P$ & polarization of electric field fluctuation in $xy$-plane \\ \hline %
$+$, $-$ & left- and right-handed electric field polarizations in $xy$-plane \\ \hline %
$\mathbf{e}_E$, & unimodular electric field polarization, and  \\ %
$\mathbf{e}_E^+$, $\mathbf{e}_E^-$ & its left- and right-handed components \\ \hline %
$\mathbf{k}$, $k$, $e_k$ & wave vector ($k=|\,\mathbf{k}|$) and its unimodular vector ($e_k = \mathbf{k} / k$) \\ \hline %
$k_\parallel$, $k_\perp$ & parallel and perpendicular components of $\mathbf{k}$ \\ \hline %
$\theta$ & angle between the propagation and the ambient magnetic field \\  %
& $\theta = \arctan (k_\perp / k_\parallel)$ \\ \hline %
$j$ ($\alpha$, $\rm p$, $\rm e$) &  plasma species $j$ in an
{\rm alpha}/{\rm proton}/{\rm electron} ($\alpha/{\rm p}/{\rm e}$) plasma \\ \hline %
$m_j$ ($m_\alpha$, $m_{\rm p}$, $m_{\rm e}$) & mass of species $j$ \\ \hline %
$z_j$ ($z_\alpha$, $z_{\rm p}$, $z_{\rm e}$) & electric charge number of species $j$ \, ($z_\alpha= 2$, \, $z_{\rm p} = z_{\rm e} = 1$) \\ \hline %
$n_j$ ($n_\alpha$, $n_{\rm p}$, $n_{\rm e}$) & abundance of species $j$ \\  %
& (neutral charge: $n_\alpha \, z_\alpha + n_{\rm p} \, z_{\rm p} + n_{\rm e} \, z_{\rm e} =0 $)  \\ %
& (minor heavy ion:  $n_\alpha \ll n_{\rm p}$) \\ \hline %
$\omega_{{\rm p}j}$ \, ($\omega_{\rm p\alpha}$, $\omega_{\rm pp}$,
$\omega_{\rm pe}$) & plasma frequency $\omega_{{\rm p}j} = \sqrt{4
\,\pi \,
n_j \, z_j^2 \,e^2 / m_j}$ \\ \hline %
$\Omega_j$ & cyclotron frequency
$\Omega_j= z_j \, e \, B_0 / (m_j \, c)$\\
($\Omega_\alpha$, $\Omega_{\rm p}$, $\Omega_{\rm e}$) & ($\Omega_{\rm p} = 2 \, \Omega_\alpha$) \\ \hline %
$\omega$, $\omega_{\rm r}$, $\omega_{\rm i}$ & complex wave frequency as well as its real and imaginary parts \\ \hline %
$v_{\rm A}$ & Alfv\'{e}n speed $v_{\rm A} = B_0 / \sqrt{4 \, \pi \, n_{\rm p} \, m_{\rm p}}$ \\ \hline %
$U_{\parallel j}$ & differential speed of species $j$ relative to protons along $\mathbf{B_0}$ \\ %
($U_{\parallel \alpha}$, $U_{\parallel {\rm e}}$) &  (zero current: $n_\alpha \, z_\alpha \, U_{\parallel \alpha} + n_{\rm e} \, z_{\rm e} \, U_{\parallel {\rm e}}=0 $) \\ \hline %
$T_{\parallel j}$ ($T_{\parallel \alpha}$, $T_{\parallel {\rm
p}}$,
$T_{\parallel {\rm e}}$) & parallel temperature of species $j$  \\ \hline %
$T_{\perp j}$ ($T_{\perp \alpha}$, $T_{\perp {\rm p}}$, $T_{\perp
{\rm e}}$) &
perpendicular temperature of species $j$  \\ \hline %
$\gamma_j$ \,($\gamma_\alpha$, $\gamma_{\rm p}$, $\gamma_{\rm e}$)
&
temperature anisotropy $\gamma_j = T_{\perp j} / \, T_{\parallel j}$ \\ \hline %
$\beta_j$ & species beta defined as \\  %
($\beta_\alpha$, $\beta_{\rm p}$, $\beta_{\rm e}$)&  a ratio of species $j$'s thermal pressure to magnetic pressure \\ %
& $\beta_j = 8 \, \pi \, n_j \, k_{\rm B} \, T_{\parallel j} / B_0^2$ \\ \hline %
$\beta$& plasma beta ($\beta = \beta_\alpha + \beta_{\rm p} + \beta_{\rm e}$) \\ \hline %
\end{tabular}
\end{table}

\newpage
\begin{table}
\begin{tabular}{ll}\hline %
$v_{\parallel j}$ \,($v_{\parallel \alpha}$, $v_{\parallel {\rm p}}$, $v_{\parallel {\rm e}}$) & parallel thermal speed $v_{\parallel j} = \sqrt{2 \, k_{\rm B} T_{\parallel j} / m_j}$ \\ \hline %
$v_{\perp j}$ \,($v_{\perp \alpha}$, $v_{\perp {\rm p}}$, $v_{\perp {\rm e}}$) & perpendicular thermal speed $ v_{\perp j} = \sqrt{2 \, k_{\rm B} T_{\perp j} / m_j}$ \\ \hline %
$f_j$ \, ($f_\alpha$, $f_{\rm p}$, $f_{\rm e}$) & gyrotropic velocity distribution function of species $j$ \\ %
& in terms of $v_{\parallel j}$ and $v_{\perp j}$ \\ \hline %
$\xi_j$ \,($\xi_\alpha$, $\xi_{\rm p}$, $\xi_{\rm e}$) & arbitrary-order general resonance factor of species $j$ \\ \hline %
$\xi^0_j$ \,($\xi^0_\alpha$, $\xi^0_{\rm p}$, $\xi^0_{\rm e}$) & zero-order Landau resonance factor of species $j$ \\ \hline %
$\xi^+_j$ \,($\xi^+_\alpha$, $\xi^+_{\rm p}$, $\xi^+_{\rm e}$) & first-order cyclotron resonance factor of species $j$ \\  \hline %
$a_{\parallel j}$ \,($a_{\parallel \alpha}$, $a_{\parallel {\rm p}}$, $a_{\parallel {\rm e}}$) & normalized parallel acceleration rate of species $j$\\ \hline %
$Q_{\parallel j}$ \,($Q_{\parallel \alpha}$, $Q_{\parallel {\rm p}}$, $Q_{\parallel {\rm e}}$) & normalized parallel heating rate of species $j$ \\ \hline %
$Q_{\perp j}$ \, ($Q_{\perp \alpha}$, $Q_{\perp {\rm p}}$) & normalized perpendicular heating rate of specie $j$ \\ \hline %
\end{tabular}
\end{table}

\newpage
\begin{table}
\caption{Fixed prescription among all initial plasma
parameters}\label{Tab2}
\begin{tabular}{ll}
\hline Symbol & Value \\ \hline %
parallel temperature ratio & $4:1:1$ \\ %
$T_{\parallel \alpha}: T_{\parallel {\rm p}}: T_{\parallel {\rm e}}$ & \\ \hline %
temperature anisotropy of alpha particles $\gamma_\alpha$ & 1 \\ \hline %
temperature anisotropy of protons $\gamma_{\rm p}$ & 1 \\ \hline %
temperature anisotropy of electrons $\gamma_{\rm e}$ & 1 \\ \hline %
\end{tabular}
\end{table}

\begin{table}
\caption{Parametric study of initial plasma
parameters}\label{Tab3}
\begin{tabular}{llllll} \hline %
Section & Figure & Propagation & Density ratio & Differential & Proton beta \\ %
&& angle $\theta$ & $n_\alpha$:$n_{\rm p}$ & speed of alpha  & $\beta_{\rm p}$  \\%
&&&& particles $U_{\parallel \alpha}$ & \\ \hline %
3, 4  & 1 -- 8 & $0^\circ$ -- $89^\circ$ & 0.02 & 0 & 0.01 \\  \hline %
5 & --- & $0^\circ$ -- $89^\circ$ & 0 & 0 & 0.01 \\ \hline %
6 & 9, 10 & $60^\circ$ & 0.02 & 0 -- 1 $v_{\rm A}$ & 0.01 \\ \hline %
7 & --- & $60^\circ$ & 0.02 & 0 & 0.001 -- 10 \\ \hline %
\end{tabular}
\end{table}

\clearpage

\newpage

\newpage

\newpage

\end{article}

\begin{thebibliography}{18}
\providecommand{\natexlab}[1]{#1} \expandafter\ifx\csname
urlstyle\endcsname\relax
  \providecommand{\doi}[1]{doi:\discretionary{}{}{}#1}\else
  \providecommand{\doi}{doi:\discretionary{}{}{}\begingroup
  \urlstyle{rm}\Url}\fi
\bibitem[\protect\citeauthoryear{Axford and McKenzie}{1992}]{Axford1992}
Axford, W.~I., McKenzie, J.~F.: 1992, The origin of high speed
solar wind streams. In: Marsch, E., Schwenn, R. (eds.) {\it Solar
Wind Seven}, {\it COSPAR Colloq.} {\bf 3}, Pergamon Press, 1--5.

\bibitem[\protect\citeauthoryear{Bale {\it et al.}}{2005}]{Bale2005}
Bale, S.~D., Kellogg, P.~J., Mozer, F.~S., Horbury, T.~S., Reme,
H.: 2005, Measurement of the electric fluctuation spectrum of
magnetohydrodynamic turbulence. {\it Phys. Rev. Lett.} {\bf 94},
215002.

\bibitem[\protect\citeauthoryear{Barnes and Hollweg}{1974}]{Barnes1974}
Barnes, A., Hollweg, J.~V.: 1974, Large-amplitude hydromagnetic
waves. {\it J. Geophys. Res.} {\bf 79}, 2302--2318.

\bibitem[\protect\citeauthoryear{Chandran}{2008}]{Chandran2008}
Chandran, B.~D.: 2008, Weakly turbulent magnetohydrodynamic waves
in compressible low-$\beta$ plasmas. {\it Phys. Rev. Lett.} {\bf
101}, 269903.

\bibitem[\protect\citeauthoryear{Cole}{1978}]{Cole1978}
Coles, W.~A.: 1978, Interplanetary scintillation. {\it Space Sci.
Rev.} {\bf 21}, 411--425.

\bibitem[\protect\citeauthoryear{Cranmer {\it et al.}}{1999}]{Cranmer1999}
Cranmer, S.~R., Kohl, J.~L., Noci, G., Antonucci, E., Tondello,
G., Huber, M.~C.~E., {\it et al.}: 1999, An empirical model of a
polar coronal hole at solar minimum. {\it Astrophys. J.} {\bf
511}, 481--501.

\bibitem[\protect\citeauthoryear{Cranmer and Ballegooijen}{2003}]{Cranmer2003}
Cranmer, S.~R., van Ballegooijen, A.~A.: 2003, Alfv\'{e}nic
turbulence in the extended solar corona: Kinetic effects and
proton heating. {\it Astrophys. J.} {\bf 594}, 573--591.

\bibitem[\protect\citeauthoryear{Dasso {\it et al.}}{2005}]{Dasso2005}
Dasso, S., Milano, L.~J., Matthaeus, W.~H., Smith, C.~W.: 2005,
Anisotropy in fast and slow solar wind fluctuations. {\it
Astrophys. J. Lett.} {\bf 635}, L181--L184.

\bibitem[\protect\citeauthoryear{Gary}{1993}]{Gary1993}
Gary, S.~P.: 1993, {\it Theory of Space Plasma
Microinstabilities}. Cambridge University Press, UK.

\bibitem[\protect\citeauthoryear{Gary and Nishimura}{2004}]{Gary2004a}
Gary, S.~P., Nishimura, K.: 2004, Kinetic {Alfv\'{e}n} waves:
{Linear} theory and a particle-in-cell simulation. {\it J.
Geophys. Res.} {\bf 109}, A02109.

\bibitem[\protect\citeauthoryear{Gary and Borovsky}{2004}]{Gary2004b}
Gary, S.~P., Borovsky, J.~E.: 2004, {Alfv\'{e}n}-cyclotron
fluctuations: {Linear Vlasov} theory. {\it J. Geophys. Res.} {\bf
109}, A06105.

\bibitem[\protect\citeauthoryear{Gary and Smith}{2009}]{Gary2009}
Gary, S.~P., Smith, W.: 2009, Short-wavelength turbulence in the
solar wind: {Linear} theory of whistler and kinetic {Alfv\'{e}n}
fluctuations. {\it J. Geophys. Res.} {\bf 114}, A12105.

\bibitem[\protect\citeauthoryear{Gogoberidze {\it et al.}}{2009}]{Gogoberidze2009}
Gogoberidze G., Mahajan, S.~M., Poedts, S.: 2009, Weak and strong
regimes of incompressible magnetohydrodynamic turbulence. {\it
Phys. Plasmas} {\bf 16}, 072304.

\bibitem[\protect\citeauthoryear{Goldreich and Sridhar}{1995}]{Goldreich1995}
Goldreich, P., Sridhar, S.: 1995, Toward a theory of interstellar
turbulence. II: strong {Alfv\'{e}n} turbulence. {\it Astrophys.
J.} {\bf 438}, 763--775.

\bibitem[\protect\citeauthoryear{Goldstein {\it et al.}}{2000}]{Goldstein2000}
Goldstein, B.~E., Neugebauer, M., Zhang, L.~D., Gary, S.~P.: 2000,
Observed constraint on proton-proton relative velocities in the
solar wind. {\it Geophys. Res. Lett.} {\bf 27}, 53--56.

\bibitem[\protect\citeauthoryear{He {\it et al.}}{2011}]{He2011}
He, J., Marsch, E., Tu, C., Yao, S., Tian, H.: 2011, Possible
evidence of Alfv\'{e}n-cyclotron waves in the angle distribution
of magnetic helicity of solar wind turbulence. {\it Astrophys. J.}
{\bf 731}, 85.

\bibitem[\protect\citeauthoryear{Hollweg and Isenberg}{2002}]{Hollweg2002b}
Hollweg, J.~V., Isenberg, P.~A.: 2002, Generation of the fast
solar wind: a review with emphasis on the resonant cyclotron
interaction. {\it J. Geophys. Res.} {\bf 107}, 1147.

\bibitem[\protect\citeauthoryear{Horbury {\it et al.}}{2005}]{Horbury2005}
Horbury, T.~S., Forman, M. A., Oughton, S.: 2005, Spacecraft
observations of solar wind turbulence: an overview. {\it J. Plasma
Phys. Controlled Fusion} {\bf 47}, B703.

\bibitem[\protect\citeauthoryear{Isenberg}{2001}]{Isenberg2001}
Isenberg, P.~A.: 2001, The kinetic shell model of coronal heating
and acceleration by ion cyclotron waves 2: {Inward} and outward
propagating waves. {\it J. Geophys. Res.} {\bf 106}, 29249--29260.

\bibitem[\protect\citeauthoryear{Kohl {\it et al.}}{1995}]{Kohl1995}
Kohl, J.~L., Esser, R., Gardner, L.~D., Habbal, S., Daigneau,
P.~S., Dennis, E.~F., {\it et al.}: 1995, The Ultraviolet
Coronagraph Spectrometer for the Solar and Heliospheric
Observatory. {\it Solar Phys.} {\bf 162}, 313--356.

\bibitem[\protect\citeauthoryear{Kohl {\it et al.}}{1998}]{Kohl1998}
Kohl, J.~L., Noci, G., Antonucci, E., Tondello, G., Huber,
M.~C.~E., Cranmer, S.~R., {\it et al.}: 1998, {UVCS/SOHO}
empirical determinations of anisotropic velocity distributions in
the solar corona. {\it Astrophys. J.} {\bf 501}, L127--L131.

\bibitem[\protect\citeauthoryear{Leamon {\it et al.}}{1998}]{Leamon1998}
Leamon, R.~J., Smith, C.~W., Ness, N.~F., Matthaeus, W.~H., Wong,
H.~K.: 1998, Observational constraints on the dynamics of the
interplanetary magnetic field
  dissipation range. {\it J. Geophys. Res.} {\bf 103}, 4775--4787.

\bibitem[\protect\citeauthoryear{Li and Habbal}{2001}]{li2001}
Li, X., Habbal, S.~R.: 2001, Damping of fast and ion cyclotron
oblique waves in the multi-ion fast solar wind. {\it J. Geophys.
Res.} {\bf 106}(A6), 10669--10680.

\bibitem[\protect\citeauthoryear{Li {\it et al.}}{1998}]{Li1998}
Li, X., Habbal, S.~R., Kohl, J.~L., Noci, G.~C.: 1998, The effect
of temperature anisotropy on observations of Doppler dimming and
pumping in the inner corona. {\it Astrophys. J.} {\bf 501},
L133--L137.

\bibitem[\protect\citeauthoryear{Li {\it et al.}}{2010}]{Li2010b}
Li, X., Lu, Q., Chen, Y., Li, B., Xia, L.: 2010, A kinetic
{Alfv\'{e}n} wave and the proton distribution function in the fast
solar wind. {\it Astrophys. J. Lett.} {\bf 719}, L190--L193.

\bibitem[\protect\citeauthoryear{MacBride {\it et al.}}{2010}]{MacBride2010}
MacBride, B.~T., Smith, C.~W., Vasquez, B.~J.: 2010,
Inertial-range anisotropies in the solar wind from 0.3 to 1 {AU:
Helios} 1 observations. {\it J. Geophys. Res.} {\bf 115}, A07105.

\bibitem[\protect\citeauthoryear{Markovskii and Hollweg}{2004}]{Markovskii2004}
Markovskii, S.~A., Hollweg, J.~V.: 2004, Intermittent heating of
the solar corona by heat flux-generated ion cyclotron waves. {\it
Astrophys. J.} {\bf 609}, 1112--1122.

\bibitem[\protect\citeauthoryear{Markovskii {\it et al.}}{2010}]{Markovskii2010}
Markovskii, S.~A., Vasquez, B.~J., Chandran, B.~D.~G.: 2010,
Perpendicular proton heating due to energy cascade of fast
magnetosonic waves in the solar corona.
  {\it Astrophys. J.} {\bf 709}, 1003--1008.

\bibitem[\protect\citeauthoryear{Marsch and Tu}{2001}]{Marsch2001}
Marsch, E., Tu, C.~Y.: 2001, Heating and acceleration of coronal
ions interacting with plasma waves through cyclotron and {Landau}
resonance. {\it J.
  Geophys. Res.} {\bf 106}, 227--238.

\bibitem[\protect\citeauthoryear{Marsch {\it et al.}}{1982}]{Marsch1982}
Marsch, E., Rosenbauer, H., Schwenn, R., Muehlhaeuser, K.~H.,
Neubauer, F.~M.: 1982, Solar wind helium ions - observations of
the {Helios} solar probes between 0.3 and 1 {AU}. {\it J. Geophys.
Res.} {\bf 87}, 35--51.

\bibitem[\protect\citeauthoryear{Matthaeus {\it et al.}}{1990}]{Matthaeus1990}
Matthaeus, W.~H., Goldstein, M.~L., Roberts, D.~A.: 1990, Evidence
for the presence of quasi-two-dimensional nearly incompressible
fluctuations in the solar wind. {\it J. Geophys. Res.} {\bf 95},
20673--20683.

\bibitem[\protect\citeauthoryear{Neugebauer and Snyder}{1962}]{Neugebauer1962}
Neugebauer, M., Snyder, C.~W.: 1962, Solar plasma experiment. {\it
Science} {\bf 138}, 1095--1097.

\bibitem[\protect\citeauthoryear{Neugebauer {\it et al.}}{1996}]{Neugebauer1996}
Neugebauer, M., Goldstein, B.~E., Smith, E.~J., Feldman, W.~C.:
1996, Ulysses observations of differential alpha-proton streaming
in the solar wind. {\it J. Geophys. Res.} {\bf 101}, 17047--17055.

\bibitem[\protect\citeauthoryear{Osmane {\it et al.}}{2010}]{Osmane2010}
Osmane, A., Hamza, A.~M., Meziane, K.: 2010, On the generation of
proton beams in fast solar wind in the presence of obliquely
propagating {Alfv\'{e}n} waves. {\it J. Geophys. Res.} {\bf 115},
A05101.

\bibitem[\protect\citeauthoryear{Parker}{1958}]{Parker1958}
Parker, E.~N.: 1958, Dynamics of the interplanetary gas and
magnetic fields. {\it Astrophys. J.} {\bf 128}, 664--676.

\bibitem[\protect\citeauthoryear{Perez and Boldyrev}{2008}]{Perez2008}
Perez, J. C., Boldyrev, S.: 2008, On weak and strong
magnetohydrodynamic turbulence. {\it Astrophys. J. Lett.} {\bf
672}, L61--L64.

\bibitem[\protect\citeauthoryear{Podesta {\it et al.}}{2010}]{Podesta2010}
Podesta, J.~J., Borovsky, J.~E., Gary, S.~P.: 2010, A kinetic
{Alfv\'{e}n} wave cascade subject to collisionless damping cannot
reach electron scales in the solar wind at 1 {AU}. {\it Astrophys.
J.} {\bf 712}, 685.

\bibitem[\protect\citeauthoryear{Sahraoui {\it et al.}}{2009}]{Sahraoui2009}
Sahraoui, F., Goldstein, M.~L., Robert, P., Khotyaintsev, Y.~V.:
2009, Evidence of a cascade and dissipation of solar-wind
turbulence at the electron gyroscale. {\it Phys. Rev. Lett.} {\bf
102}, 231102.

\bibitem[\protect\citeauthoryear{Sahraoui {\it et al.}}{2010}]{Sahraoui2010}
Sahraoui, F., Goldstein, M.~L., Belmont, G., Canu, P., Rezeau, L.:
2010, Three dimensional anisotropic $k$ spectra of turbulence at
subproton scales in the solar wind. {\it Phys. Rev. Lett.} {\bf
105}, 131101.

\bibitem[\protect\citeauthoryear{Smith {\it et al.}}{2001}]{Smith2001}
Smith, C.~W., Mullan, D.~J., Ness, N.~F., Skoug, R.~M., Steinberg,
J.: 2001, Day the solar wind almost disappeared: Magnetic field
fluctuations wave refraction and dissipation. {\it J. Geophys.
Res.} {\bf 106}, 18625--18634.

\bibitem[\protect\citeauthoryear{Stix}{1992}]{Stix1992}
Stix, T.~H.: 1992, \textit{Waves in Plasmas}. Springer, 237--304.

\bibitem[\protect\citeauthoryear{Wick {\it et al.}}{2011}]{Wick2011}
Wicks, R.~T., Horbury, T.~S., Chen, C.~H.~K., Schekochihin, A.~A.:
2011, Anisotropy of imbalanced {Alfv\'{e}n} turbulence in fast
solar wind. {\it Phys. Rev. Lett.} {\bf 106}, 045001.

\end{thebibliography}
\end{document}